\documentclass[12pt]{article}
\pdfoutput=1
\usepackage{amssymb, amsmath,amsfonts}
\usepackage{multirow}
\usepackage{mathrsfs}
\usepackage{array}
\usepackage{cite}
\usepackage{booktabs}
\usepackage{tikz}
\usepackage{pgfplots}
\usepackage{float}
\usepackage{graphicx}
\usepackage[pdftex, bookmarks=true,colorlinks,linkcolor=red,urlcolor=blue,citecolor=blue]{hyperref}

\makeatletter\@addtoreset{equation}{section}\makeatother

\def\be{\begin{equation}}
\def\ee{\end{equation}}
\def\bea{\begin{eqnarray}}
\def\eea{\end{eqnarray}}

\newcommand{\nn}{\nonumber}

\def\Dslash{\,\,{\raise.15ex\hbox{/}\mkern-12mu D}}
\def\Dbarslash{\,\,{\raise.15ex\hbox{/}\mkern-12mu {\bar D}}}
\def\delslash{\,\,{\raise.15ex\hbox{/}\mkern-9mu \partial}}
\def\delbarslash{\,\,{\raise.15ex\hbox{/}\mkern-9mu {\bar\partial}}}
\def\pslash{\,\,{\raise.15ex\hbox{/}\mkern-9mu p}}
\def\calDslash{\,\,{\raise.15ex\hbox{/}\mkern-12mu {\cal D}}}

\makeatletter\@addtoreset{equation}{section}\makeatother

\newcommand{\preprint}[1]{\begin{table}[t]  
             \begin{flushright}               
             {#1}                             
             \end{flushright}                 
             \end{table}}                     
\renewcommand{\title}[1]{\vbox{\center\LARGE{#1}}\vspace{5mm}}
\renewcommand{\author}[1]{\vbox{\center#1}\vspace{5mm}}
\newcommand{\address}[1]{\vbox{\center\em#1}}

\def\arXiv#1{\href{http://arxiv.org/abs/#1}{arXiv:#1}}
\def\arXiv#1#2{\href{http://arxiv.org/abs/#1}{arXiv:#1}}

\def\doi#1#2{\href{http://doi.org/#1}{#2}}

\begin{document}

\unitlength = .8mm

\begin{titlepage}
\vspace{.5cm}
\preprint{}
 
\begin{center}
\hfill \\
\hfill \\
\vskip 1cm

\title{An improved holographic nodal line semimetal}
\vskip 0.5cm
{Yan Liu}\footnote{Email: {\tt yanliu@buaa.edu.cn}} and
 {Xin-Meng Wu}\footnote{Email: {\tt wu\_xm@buaa.edu.cn}} 

\address{Center for Gravitational Physics, Department of Space Science, \\ and International Research Institute
of Multidisciplinary Science,
\\ Beihang University,  Beijing 100191, China}

\end{center}
\vskip 1.5cm

\abstract{We study an improved holographic model for the strongly coupled nodal line semimetal which satisfies the duality relation between the rank two tensor operators $\bar{\psi}\gamma^{\mu \nu}\psi$ and $\bar{\psi}\gamma^{\mu \nu}\gamma^5\psi$}. 
We introduce a Chern-Simons term and a mass term in the  bulk for a complex two form field which is dual to the above tensor operators 
and 
the duality relation is automatically satisfied from holography. 
We find that there exists a quantum phase transition from a topological nodal line semimetal phase to a trivial phase. In the topological phase, there exist multiple nodal lines in the fermionic spectrum which are topologically nontrivial. 
The bulk geometries are different from the previous model without the duality constraint, while the resulting  properties are qualitatively similar to those in that model.   
This improved model provides a more natural ground  
 to analyze transports or other properties of strongly coupled nodal line semimetals.

\vfill

\end{titlepage}

\begingroup 
\hypersetup{linkcolor=black}
\tableofcontents
\endgroup

\section{Introduction}

Semimetals are critical states in the phase transition between insulators and conductors. In general, the Fermi surfaces in semimetals are of zero area and could be discrete points (Dirac semimetal or Weyl semimetal, i.e. DSM or WSM) or nodal lines (nodal line semimetal, i.e. NLSM). In topological semimetals these points or nodal lines are stable and could not be removed by perturbations of the system without breaking certain symmetries.\footnote{
The nodal points or nodal lines in semimetals are the sources of the Berry curvature and give rise to topological charges \cite{Burkov:2017, RMP}. } 
Therefore topological semimetals exhibit lots of robust and exotic quantum properties and have attracted enormous  research interests during the past few years \cite{RMP, rev1}. 

Topological semimetals  are beyond the conventional Landau-Ginzburg paradigm and are characterized by the topological properties of the wave functions of the system. 
Most of the current models for topological semimetals are constructed based on the existence of quasiparticles, where one starts from a weakly coupled band theory of 
an effective Hamiltonian. An important and challenging question is whether and how the topological structures  change if the system is strongly coupled. Without a clear quasiparticle description in the strongly coupled system \cite{book}, would nontrivial topological states still exist and if yes how could we describe them? 

The holographic duality is a useful technique in describing strongly coupled systems in quantum physics by mapping them to simple gravitational problems \cite{book, book0, review}. Holography has been applied to the study of the topological nature in Weyl semimetals \cite{Landsteiner:2015lsa} and nodal line semimetals \cite{Liu:2018bye} which uncovered lots of novel properties of topological semimetals at strong coupling \cite{Landsteiner:2015pdh, Landsteiner:2016stv, Copetti:2016ewq, Grignani:2016wyz, Ammon:2016mwa, Ammon:2018wzb, Baggioli:2018afg, Liu:2018spp, Song:2019asj, Ji:2019pxx, Tanaka:2020yax, Juricic:2020sgg, Baggioli:2020cld, Kiczek:2020qsw, Gursoy:2012ie, fadafan}. See \cite{Landsteiner:2019kxb} for a recent review on this topic. 

The holographic nodal line semimetals were first studied in  \cite{Liu:2018bye} and the key ingredients are a source for a rank two operator $\bar{\psi}\gamma^{\mu \nu}\psi$ and a mass deformation parameter. In \cite{Liu:2018bye} a complex two form field is introduced in the bulk whose real part is dual to the above operator $\bar{\psi}\gamma^{\mu \nu}\psi$ while the imaginary part is dual to the other 
rank two operator  
$\bar{\psi}\gamma^{\mu \nu}\gamma^5\psi$. These two operators $\bar{\psi}\gamma^{\mu \nu}\psi$ and $\bar{\psi}\gamma^{\mu \nu}\gamma^5\psi$ on the field theory are not independent and we have the duality relation 
$
\bar{\psi}\gamma^{\mu \nu}\gamma^5\psi=-\frac{i}{2}\varepsilon^{\mu\nu}_{~~\alpha\beta}\bar{\psi}\gamma^{\alpha \beta}\psi\,\,.
$
However, this duality relation has not been considered in the  holographic model of \cite{Liu:2018bye}, where it is apparent that the two form field in the bulk is not exactly dual to the operators $\bar{\psi}\gamma^{\mu \nu}\psi$ and $\bar{\psi}\gamma^{\mu \nu}\gamma^5\psi$. In weakly coupled nodal line semimetals, the two operators $\bar{\psi}\gamma^{\mu \nu}\psi$ and $\bar{\psi}\gamma^{\mu \nu}\gamma^5\psi$ above are crucial to the related physics of nodal line semimetals. Therefore in holography it is more natural to take a two form field in the bulk to be exactly dual to these rank two operators in order to sharply describe the physics of strongly nodal line semimetals, as the holographic model in \cite{Liu:2018bye} might include the physics of other rank two operators which is not related to the featured physics of nodal line semimetals. Thus we should consider an improved holographic nodal line semimetal model in which the duality relation of the rank two operators are indicated.\footnote{We thank Carlos Hoyos and Elias Kiritsis for helpful discussion on this point.} A similar issue has been studied in the framework of AdS/QCD in \cite{Alvares:2011wb} where a  rank two field corresponds to vector mesons satisfying the self-dual constraint and it turns out that the bulk action should be a first order Chern-Simons term with a mass term. The duality of the operators on the boundary field theory automatically follows from the bulk equation of motion.  

Our strategy is to follow the constructions in \cite{Alvares:2011wb} to improve the holographic nodal line semimetal of \cite{Liu:2018bye} to include the self-duality condition to make the holographic theory more natural. 
We shall focus on the zero temperature physics at which the topological properties are most manifest, by tuning  
the source of the rank two operators and the mass deformation. 
The bulk IR geometries vary with different emergent low energy symmetries. We also investigate the fermionic spectral function of this improved holographic model and multiple nodal lines of the spectral function are found. The topological invariants of these nodal lines are also studied. From the field theoretical point of view a quantum phase transition is triggered by the strength of external sources, which is confirmed by studying the fermionic spectral functions. 
It is not obvious whether the duality constraint might induce any different effects on the holographic model, e.g. whether the order of the quantum phase transition might change, whether the topological state still exist and if 
yes then what are the corresponding topological invariants. 
We find that the properties in this improved holographic nodal line semimetal model share lots of similar physics as the one in \cite{Liu:2018bye}, i.e. there is no qualitative change. Therefore, our improved model serves as a natural holographic model for strongly coupled NLSM. 

The organization of this paper is as follows. In Sec. \ref{sec2}, we construct the improved holographic nodal line semimetal by including the self-duality condition. Zero temperature geometries of the system are studied to uncover the phase structures and the topological properties.  
In Sec. \ref{sec:arpes} we study the fermionic spectral function to uncover the Fermi surface of the dual field theory. In Sec. 
\ref{sec:ti} we study the topological invariants of each nodal line of the fermionic spectal function. We conclude with discussions in Sec. \ref{sec:cd}. Some details of calculations are collected in four appendices. 

\section{An improved holographic nodal line semimetal}
\label{sec2}
In this section we shall first analyze the field theory of nodal line semimetals and point out that we should consider the duality of rank two operators in the holographic construction. 
Then we follow \cite{Alvares:2011wb} to improve the holographic nodal line semimetal of \cite{Liu:2018bye} and study the zero temperature solutions. 

\subsection{Field model}

Topological nodal line semimetal 
is realized in a Lorentz violating field theoretical model 
\cite{burkov1, rev1, Liu:2018bye}
with the following Lagrangian\footnote{Note that we work in the Minkowski metric with most plus convention. The gamma matrices are the same as in appendix \ref{appD}.} 
\be
\mathcal{L}=\bar{\psi}\big(\gamma^\mu\partial_\mu-m-\gamma^{\mu\nu}b_{\mu\nu}\big)\psi\,,
\label{eqL:previous}
\ee
where $\bar{\psi}=\psi^{\dagger}i\gamma^0$, $\gamma^{\mu\nu}=\frac{i}{2}[\gamma^\mu, \gamma^\nu]$ and $b_{\mu \nu}=-b_{\nu \mu}$ is an anti-symmetric two form field. If we turn on real $b_{xy}$, this system describes a nodal line semimetal. The equation of motion for the Dirac fermion is
\be
\big(\gamma^\mu \partial_\mu-m-\gamma^{\mu\nu}b_{\mu\nu}\big)\psi=0\,.
\ee
The Hamiltonian matrix can be obtained by writing this equation as a Schrodinger equation
\be
i\frac{\partial \psi}{\partial t }=-i\gamma^0\big(\gamma^i\partial_i-m-\gamma^{\mu\nu}b_{\mu\nu}\big)\psi\equiv \hat{H}\psi\,.
\ee
Therefore, the band structure and the eigenstates of this Dirac system can be determined and reveal a quantum phase transition from the nodal line semimetal phase to gapped system by tuning the ratio between $b_{xy}$ and $m$. 

\begin{figure}[h!]
  \centering
\includegraphics[width=0.45\textwidth]{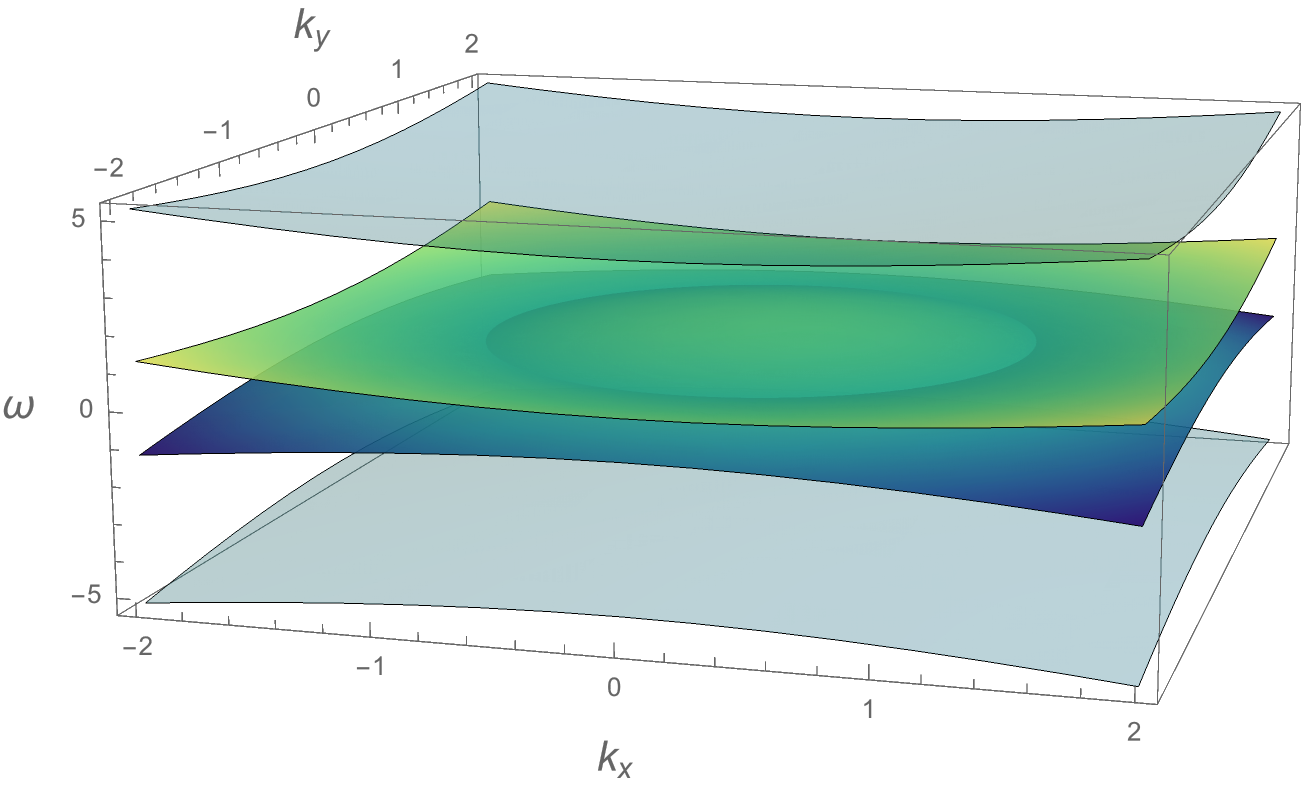}
\hspace{10mm}
\includegraphics[width=0.45\textwidth]{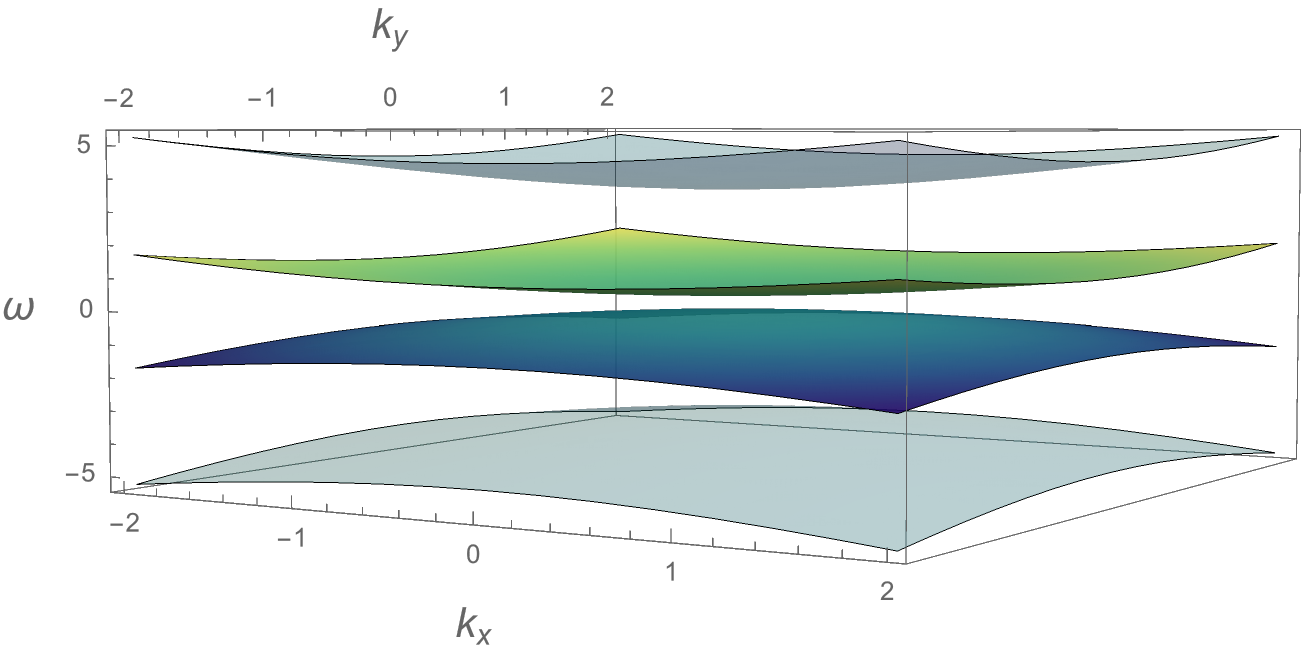}
  \caption{\small The illustration of the band structures with $k_z=0$ for the nodal line semimetal phase (left) and the gapped phase (right). In the nodal line semimetal phase, the crossing band forms a nodal circle where the excitations can be described by the Weyl fermions. There is no band crossing at any nonzero $k_z$. }
  \label{fig:weakband}
\end{figure}

However, in the four-dimensional Minkowski spacetime, the two anti-symmetric tensor operators $\bar{\psi}\gamma^{\mu \nu}\psi$ and $\bar{\psi}\gamma^{\mu \nu}\gamma^5\psi$ are not independent and are related by the duality relation 
\be
\label{eq:selfduality}
\bar{\psi}\gamma^{\mu \nu}\gamma^5\psi=-\frac{i}{2}\varepsilon^{\mu\nu}_{~~\alpha\beta}\bar{\psi}\gamma^{\alpha \beta}\psi\,,
\ee
where $\varepsilon_{txyz}=1$, $\bar{\psi}\gamma^{\mu \nu}\psi$ is a pure real  tensor operator while $\bar{\psi}\gamma^{\mu \nu}\gamma^5\psi$ is a pure imaginary tensor operator. 
As a consequence, it is more natural to take the 
operator $\bar{\psi}\gamma^{\mu \nu}\gamma^5\psi$ into consideration and the action (\ref{eqL:previous}) is modified into 
\be
\mathcal{L}=\bar{\psi}\big(\gamma^\mu\partial_\mu-m-\gamma^{\mu\nu}b_{\mu\nu}+\gamma^{\mu\nu}\gamma^5b^5_{\mu\nu}\big)\psi\,.
\label{eqL:imNLSM}
\ee
Due to the duality relation of the two form operators  (\ref{eq:selfduality}), 
we turn on $b_{xy}$, $b_{yx}=-b_{xy}$ and their dual part of  $b^5_{\mu\nu}$, i.e., $b^5_{tz}=-b^5_{zt}=i b_{xy}$. 
Here the external source $b^5_{\mu\nu}$ is set to be pure imaginary to make the Hamiltonian real. 
With this choice of Lagrangian in (\ref{eqL:imNLSM}), 
the band structure, i.e. $E_{\pm}=\pm\sqrt{(4b_{xy} \pm \sqrt{m^2 + k_x^2 + k_y^2})^2 + k_z^2}$ is the same as that described with the Lagrangian of (\ref{eqL:previous}) up to a prefactor of $b_{xy}$  \cite{Liu:2018bye}. 
The band crossing appears at the $k_z=0$ plane and forms a circle with a radius $\sqrt{k_x^2 + k_y^2}=\sqrt{16b_{xy}^2 - m^2}$ when $16 b_{xy}^2-m^2>0$, while for finite $k_z$ the energy band is gapped. 
The model (\ref{eqL:imNLSM}) with the duality relation of the rank two operators (\ref{eq:selfduality}) is called the improved nodal line semimetal. 
In this model, there still exists a 
quantum phase transition from a nodal line semimetal to a gapped phase, as shown in Fig. \ref{fig:weakband}. 

\vspace{.3cm}

\subsubsection{From weakly coupled model to strongly coupled model}

Starting from the properties of weakly coupled model for nodal line semimetal  (\ref{eqL:imNLSM}), we will construct 
a holographic model for strongly coupled nodal line semimetal with the duality relation. In holography, we use a  complex anti-symmetric two form field $B_{\mu\nu}$ 
to be dual to operators $\bar{\psi}\gamma^{\mu\nu}\psi$ and $\bar{\psi}\gamma^{\mu\nu}\gamma^5\psi$. More precisely, we parameterize the complex $B_{\mu\nu}$ as $B_{\mu\nu}=\frac{1}{\sqrt{2}}\big(B_{+\mu\nu}+iB_{-\mu\nu}\big)$ with the real anti-symmetric fields $B_{+\mu\nu}$ and $B_{-\mu\nu}$ the real and imaginary part in the $B_{\mu\nu}$ field, i.e. $B_{+\mu\nu}$ and $B_{-\mu\nu}$,  duals to $\bar{\psi}\gamma^{\mu\nu}\psi$ and $\bar{\psi}\gamma^{\mu\nu}\gamma^5\psi$, respectively.

The complex two form field $B_{ab}$ in the gravitational theory is dynamical. Note that $B_{ar}$ vanishes with a radial gauge. We expect the components $B_{\mu\nu}$ to be consistent with the duality condition between two tensor operators (\ref{eq:selfduality}). 
Similar issue has been discussed in AdS/QCD and there are many progresses in imposing this self-duality relation in the gravitational theory.  In particular, 
it has been proposed 
in \cite{Alvares:2011wb, Domokos:2011dn} that the action of the two form field $B_{ab}$ should be of first order, in such a way that the four-dimensional components satisfy a complex self-duality relation. 
The Chern-Simons term of $B_{ab}$ together with the mass term has been used in an improved holographic QCD model \cite{Alvares:2011wb} to investigate the physical properties of mesons, from which the self-duality relation follows from the equation of motion directly instead of being imposed as particular boundary conditions.\footnote{A kinetic term of $B_{ab}$ in the action is also allowed to be consistent with the self-duality condition and can be chosen to vanish without loss of generality.}
In this paper, we follow this approach to construct the improved holographic nodal line semimetal model that satisfies the self-duality condition.



\subsection{Holographic model}
 The action of the improved holographic nodal line semimetal model is 
\bea
\begin{split}
S&=\int d^5x\sqrt{-g}\bigg[\frac{1}{2\kappa^2}\bigg(R+\frac{12}{L^2}\bigg)-\frac{1}{4}\mathcal{F}^2-\frac{1}{4}F^2+\frac{\alpha}{3}\epsilon^{abcde}A_a \Big(3\mathcal{F}_{bc}\mathcal{F}_{de}+F_{bc}F_{de}\Big)\\
&
-(D_a \Phi)^*(D^a\Phi)-V_1(\Phi)-\frac{1}{6\eta}\epsilon^{abcde} \Big( i B_{ab}H_{cde}^*-i B_{ab}^* H_{cde}\Big)
-V_2(B_{ab})-\lambda|\Phi|^2B_{ab}^*B^{ab}\bigg]\,,
\end{split}
\label{eq:action}
\eea
where $\mathcal{F}_{ab}=\partial_a V_b-\partial_b V_a$ is the vector gauge field strength, $F_{ab}=\partial_a A_b-\partial_b A_a$ is the axial gauge field strength,  $\epsilon^{abcde}$ is the upper indexed Levi-Civita tensor, and $D_a=\nabla_a -iq_1 A_a$ is the covariant derivative.\footnote{Note that $\epsilon_{abcde} \equiv \sqrt{-g}\varepsilon_{abcde}$, with $\varepsilon_{abcde}$ the Levi-Civita symbol and $\varepsilon_{0123r}=1$.}
Note that $\alpha$ is the Chern-Simons coupling. 
$B_{ab}$ is an antisymmetric complex two form field that duals to the two tensor operators $\bar{\psi}\gamma^{\mu\nu}\psi$ and $\bar{\psi}\gamma^{\mu\nu}\gamma^5\psi$, and 
\bea
H_{abc}&=&\partial_a B_{bc}+\partial_b B_{ca}+\partial_c B_{ab}-iq_2 A_a B_{bc}-iq_2 A_b B_{ca}-iq_2 A_c B_{ab}\,
\eea
where $q_2$ is the axial charge of the two form field. $\eta$ is the coupling strength of the two form field. 
The potential terms are chosen as
\be
V_1=m_1^2 |\Phi|^2+\frac{\lambda_1}{2} |\Phi|^4\,,~~~~
V_2=m_2^2 B^*_{ab}B^{ab}\,,
\ee
where $m_1^2$ and $m_2^2$ are the mass parameters of the scalar field and the two form field. The $\lambda$ term denotes the interaction between the scalar field and the two form field. 
It should be noticed that in the previous holographic model \cite{Liu:2018bye} the self-duality of $B_{ab}$ is absent, where a canonical kinetic term together with the potential terms were used to describe the dynamics of $B_{ab}$. 
In this improved holographic nodal line semimetal model (\ref{eq:action}) we follow the approach in an AdS/QCD model \cite{Alvares:2011wb} to use the Chern-Simons term and the mass term of $B_{ab}$ to describe the dynamics in the gravitational theory. 
Different from the AdS/QCD model \cite{Alvares:2011wb} working in the probe limit around a pure AdS$_5$ background, we concentrate on how this two form field together with a scalar field deforms the bulk geometry under continuously varying boundary sources. 
From the viewpoint of RG flow, this can be interpreted as a UV fixed point flows to an IR fixed point induced by the external sources. We will also study the properties of Fermi surfaces and topological structures of the dual filed theories. 

After a variation of the total action with respect to the gauge fields, we can obtain the dual consistent currents and they satisfy
\bea
\partial_\mu J^\mu_{\text{{con}}}&=&0\,,\nn\\
\partial_\mu J^\mu_{5\text{{con}}}&=&\lim_{r\to\infty}\sqrt{-g}\bigg(-\frac{\alpha}{3}\epsilon^{r\alpha\beta\rho\sigma}(F_{\alpha\beta}F_{\rho\sigma}+\mathcal{F}_{\alpha\beta}\mathcal{F}_{\rho\sigma})+iq_1\Big[\Phi^*(D^r\Phi)-\Phi(D^r\Phi)^*\Big]+\nn\\
&&~~~~~~~~~-\frac{q_2}{\eta}\epsilon^{r\alpha\beta\rho\sigma}B_{\alpha\beta}B_{\rho\sigma}^*\bigg)+\text{c.t.}\,.\nn
\eea
Here we have not explicitly shown the counterterm for simplicity and the above conservation can be further simplified in the radial gauge. The point is that the last two terms contribute only when the non-normalizable mode of the scalar filed or two form field is switched on and it is straightforward to see that the above identities are of the same structure of the weakly coupled theory. Thus this holographic model is expected to go beyond the weakly coupled theory to a strongly coupled nodal line semimetal model. 

\subsubsection{Zero temperature solutions}
We focus on the ground state of this system and  make the following ansatz for the zero temperature solution 
\bea
\label{eq:bg-nlsm}
\begin{split}
ds^2 &= u(-dt^2+dz^2)+\frac{dr^2}{u}+f(dx^2+dy^2)\,,\\~~~
\Phi&=\phi\,,~~~\\
B_{xy}&=-B_{yx}=\mathcal{B}_{xy}\,,\\
B_{tz}&=-B_{zt}=i\mathcal{B}_{tz}\,
\end{split}
\eea
in the coordinates $\{t, x, y, z, r\}$, where all the fields $u, f, \phi, \mathcal{B}_{xy}, \mathcal{B}_{tz}$ are real functions of the radial coordinate $r$. 
Note that the component $B_{xy}$ is real while $B_{tz}$ is pure imaginary. Substituting this ansatz into the equation of motion\footnote{The equations of motion for this system can be found in Appendix \ref{appA}.} derived from the action  (\ref{eq:action}), we write down the first order differential equations of $\mathcal{B}_{xy}$ and  $\mathcal{B}_{tz}$
\bea
\begin{split}
\mathcal{B}_{tz}'-\frac{\eta \sqrt{u}}{2f}(m_2^2+\lambda \phi^2)\mathcal{B}_{xy}&=0\,,\\
\mathcal{B}_{xy}'-\frac{\eta f}{2u^{\frac{3}{2}}}(m_2^2+\lambda \phi^2)\mathcal{B}_{tz}&=0\,.
\end{split}
\eea

We make some observations and explanations at this stage.
\begin{itemize}
  \item $\mathcal{B}_{tz}$ and $\mathcal{B}_{xy}$ are two independent functions of $r$ and cannot be set to equal. 
  \item Close to UV AdS$_5$ boundary $(r\rightarrow\infty)$, $u\rightarrow r^2$, $f\rightarrow r^2$ while $\phi\rightarrow 0$. It is the term $\eta\cdot m_2^2$ that controls the conformal dimension of $B_{ab}$. In the following, we will fix $m_1^2=-3,~m_2^2=1$ and $\eta=2$ to make the expected conformal dimensions for the operators which are dual to $\Phi$ and $B_{ab}$. For simplicity, we choose the couplings $q_1=q_2=1$, $\lambda=1$ and $\lambda_1=0.1$. 
  \item With this choice of parameters, the leading order solutions of $\mathcal{B}_{tz}$ and $\mathcal{B}_{xy}$ are the same close to UV boundary, i.e. $\mathcal{B}_{tz}=\mathcal{B}_{xy}\simeq br+\cdots$. This feature follows automatically from the dynamical equations of motion above  and therefore the self-duality of the two form field is imposed in holography. 
Note that when the self-duality constraint is automatically imposed on the source terms, the duality condition should be also on the rank two tensor operators as long as we perform the variational principle on the dual field theory correctly.\footnote{We thank Carlos Hoyos for useful discussion on this point.} 
\end{itemize}
Therefore, we impose the following boundary condition which encode the duality relation of the rank two operators 
\be
\label{eq:dictionary}
\mathop{\text{lim}}_{r\rightarrow \infty}~r\phi=M\,,~~\mathop{\text{lim}}_{r\rightarrow \infty}~r^{-1}\mathcal{B}_{tz}=\mathop{\text{lim}}_{r\rightarrow \infty}~r^{-1}\mathcal{B}_{xy}=b\,,
\ee
where $M$ plays the role of the source of the scalar operator $\bar{\psi}\psi$ while $b$ corresponds to the external source term of the tensor operators $\bar{\psi}\gamma^{\mu\nu}\psi$ and $\bar{\psi}\gamma^{\mu\nu}\gamma^5\psi$. 

In the remaining context of this section, we illuminate the existence of three phases 
when tuning the dimensionless ratio between the external source strength of the scalar and tensor operators in field theory, i.e., $M/b$.  
From the numerical study of 
the free energy we found that the quantum phase transition 
is continuous, 
although the discontinuity appears in the IR-region of the bulk, i.e. $r\rightarrow 0$ in our choice of coordinates.

\noindent \emph{Topological nodal line semimetal $(M/b<0.8597)$}

\noindent The IR geometry of topological nodal line semimetal phase behaves as 
\bea
\label{eq:bgir-topo}
\begin{split}
u&=\frac{1}{8}(11+3\sqrt{13})\, r^2\left(1+\delta u~r^{\alpha_1}\right)\,,\\
f&=\sqrt{\frac{2}{3}\sqrt{13}-2}\,b_{xy0}\, r^{\alpha}\left(1+\delta f~r^{\alpha_1}\right)\,,\\
\phi&=\phi_0 r^\beta\,,\\
\mathcal{B}_{tz}&=\frac{1}{8}\sqrt{54+15\sqrt{13}} ~r^2\left(1+\delta b_{tz} ~r^{\alpha_1}\right)\,,\\
\mathcal{B}_{xy}&=b_{xy0} \, r^\alpha\left(1+\delta b_{xy} ~r^{\alpha_1}\right)\,,
\end{split}
\eea
where 
\bea
(\alpha, \beta, \alpha_1)=(0.183, 0.228, 1.273)\,, ~~~(\delta f, \delta b_{tz}, \delta b_{xy})=(-2.616, 1.720, -0.302) ~\delta u\,.
\eea
In the IR limit $r\to 0$, $ds^2$ and $B_{ab}dx^adx^b$ are invariant under the transformation $(r^{-1}, t, z)\rightarrow c(r^{-1}, t, z)$, $(x, y)\rightarrow c^{\alpha/2}(x, y)$. This implies that there is an emergent Lifshitz-type symmetry in the deep IR region. 
We can set $b_{xy0}=1$ using the scaling symmetry in the $x$-$y$ plane, i.e. the second type of scaling symmetry in appendix \ref{appC}. 
The above emergent Lifshitz scaling symmetry 
can be used to set $\delta u=\pm 1$. 
With $\delta u=-1$ the IR geometry flows  to an asymptotic AdS$_5$ boundary. 
Therefore we can take $\phi_0$ as the shooting parameter, which generates a class of solutions with 
a single dimensionless UV parameter $M/b$. 
By continuously varying $\phi_0$ in IR, the UV geometries are  the AdS boundary with continuous $M/b$. We found that this type of IR geometry exist only when $M/b<0.8597$. The connection between this type of geometry and nodal line semimetal will be discussed in section \ref{sec:arpes}. 
\vspace{.4cm}\\
\noindent \emph{Quantum critical point $((M/b)_c\simeq 0.8597)$} 

\noindent The IR geometry of the quantum critical point is 
\bea
\label{eq:bgir-QCP}
\begin{split}
u&=u_c \,r^2(1+\delta u~r^{\beta})\,,\\
f&=f_c\,r^{\alpha_c}(1+\delta f~r^{\beta})\,,\\
\phi&=\phi_c\,(1+\delta \phi~r^\beta)\,,\\
\mathcal{B}_{tz}&=b_{tzc}\,r^2(1+\delta b_{tz}r^{\beta})\,,\\
\mathcal{B}_{xy}&=b_{xyc}\,r^{\alpha_c}(1+\delta b_{xy}r^{\beta})\,,
\end{split}
\eea
with 
\bea
(u_c, ~f_c, ~\alpha_c, ~\phi_c, ~b_{tzc})=(2.735\,,~ 0.754~b_{xyc}\,, ~0.314\,, ~0.557\,, ~1.437)\,,
\eea
 and
 \bea
 \beta=1.274\,, ~~~(\delta u, \delta f, \delta b_{tz}, \delta b_{xy})=(0.882, ~-2.151,~ 1.718, ~-0.254)\delta \phi\,.
 \eea
The Lifshitz type symmetry also emerges at the deep IR region, i.e., the geometry is invariant under the transformation $(r^{-1}, t, z)\rightarrow c(r^{-1}, t, z)$, $(x, y)\rightarrow c^{\alpha_c/2}(x, y)$ when $r\rightarrow 0$. 
Using this symmetry we can set $\delta \phi=-1$.
We can also set $b_{xyc}=1$ using the scaling symmetry in the $x$-$y$ plane.   
We obtain a unique geometry 
and in UV we have 
a special value of $M/b\simeq 0.8597$. 
An interesting observation is that, the critical value in this improved model is approximately half of that in the previous model \cite{Liu:2018bye}. 
Recall that in the field theoretical models, the improved model with $b_{xy}=1/2$ produces exactly the same band structure as the previous one in \cite{Liu:2018bye} with $b_{xy}=1$. 
The critical value $(M/b)_c$ changes correspondingly with a factor $1/2$ in two models seems to also hold in holography, although the detailed geometries in the bulk are different. 
\vspace{.4cm}\\
\noindent \emph{Topological trivial phase $(M/b>0.8597)$}

\noindent The IR geometry for the trivial phase is
\bea
\begin{split}
\label{eq:bgir-trivial}
u&=(1+\frac{3}{8 \lambda_1})\, r^2\,,\\
f&=r^2\,,\\
\phi&=\sqrt{\frac{3}{\lambda_1}}+\phi_1\, r^{2(\sqrt{\frac{3+20\lambda_1}{3+8\lambda_1}}-1)}\,,\\
\mathcal{B}_{tz}&=\left(1+\frac{3}{8 \lambda_1}\right)\, b_{1} \, r^{2\sqrt{2}\frac{3\lambda+\lambda_1}{\sqrt{\lambda_1(3+8\lambda_1)}}}\,,\\
\mathcal{B}_{xy}&=b_{1}r^{2\sqrt{2}\frac{3\lambda+\lambda_1}{\sqrt{\lambda_1(3+8\lambda_1)}}}\,.
\end{split}
\eea
We can set $b_1=1$ using the scaling symmetries and take $\phi_1$ as the shooting parameter. 
This type of IR geometry only exist for $M/b>0.8597$, when we continuously tune the shooting parameter 
$\phi_1$. 

\begin{figure}[h!]
  \centering
 \begin{minipage}[b]{1\textwidth}
 \centering
  \includegraphics[width=0.32\textwidth]{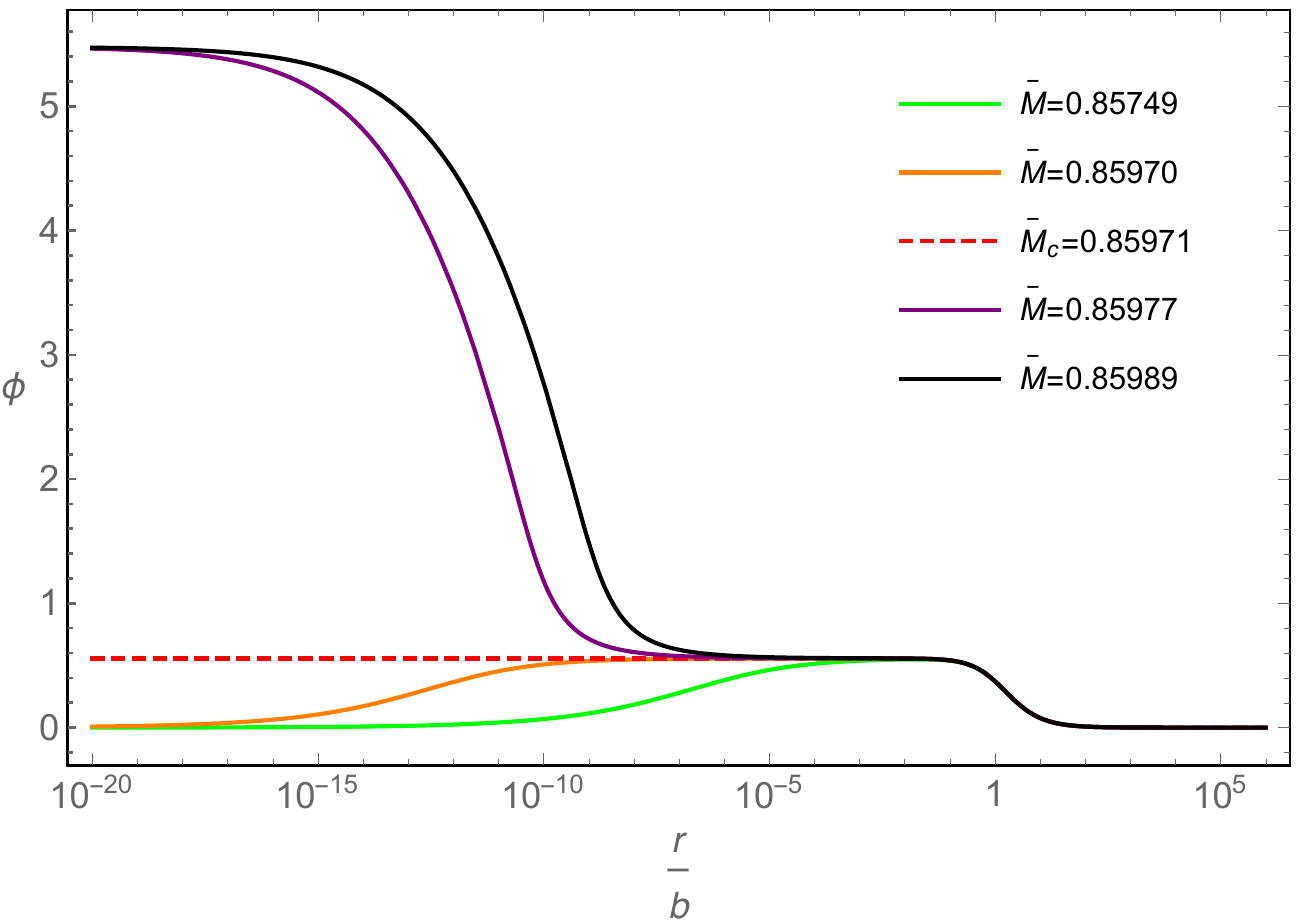} 
 \includegraphics[width=0.33\textwidth]{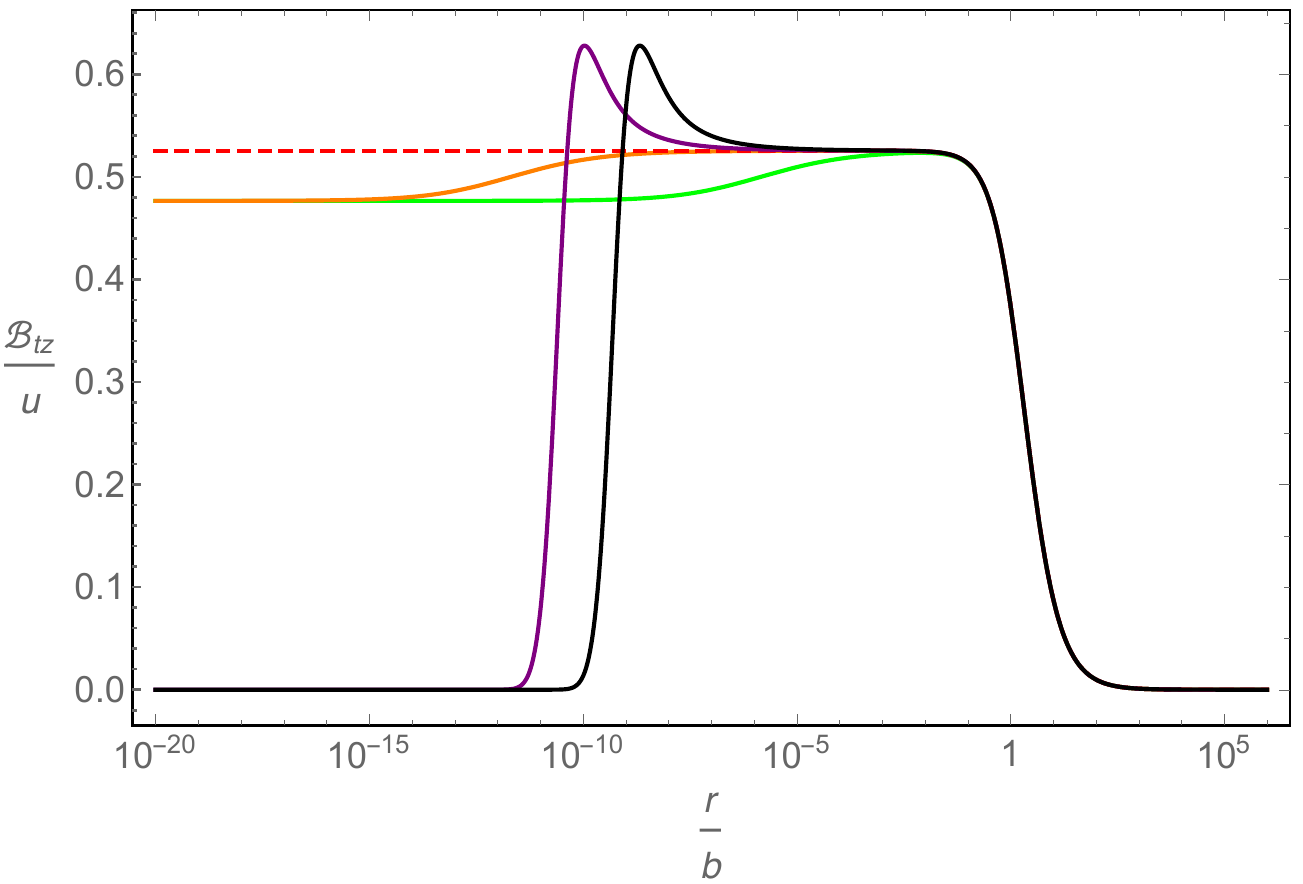}
 \includegraphics[width=0.33\textwidth]{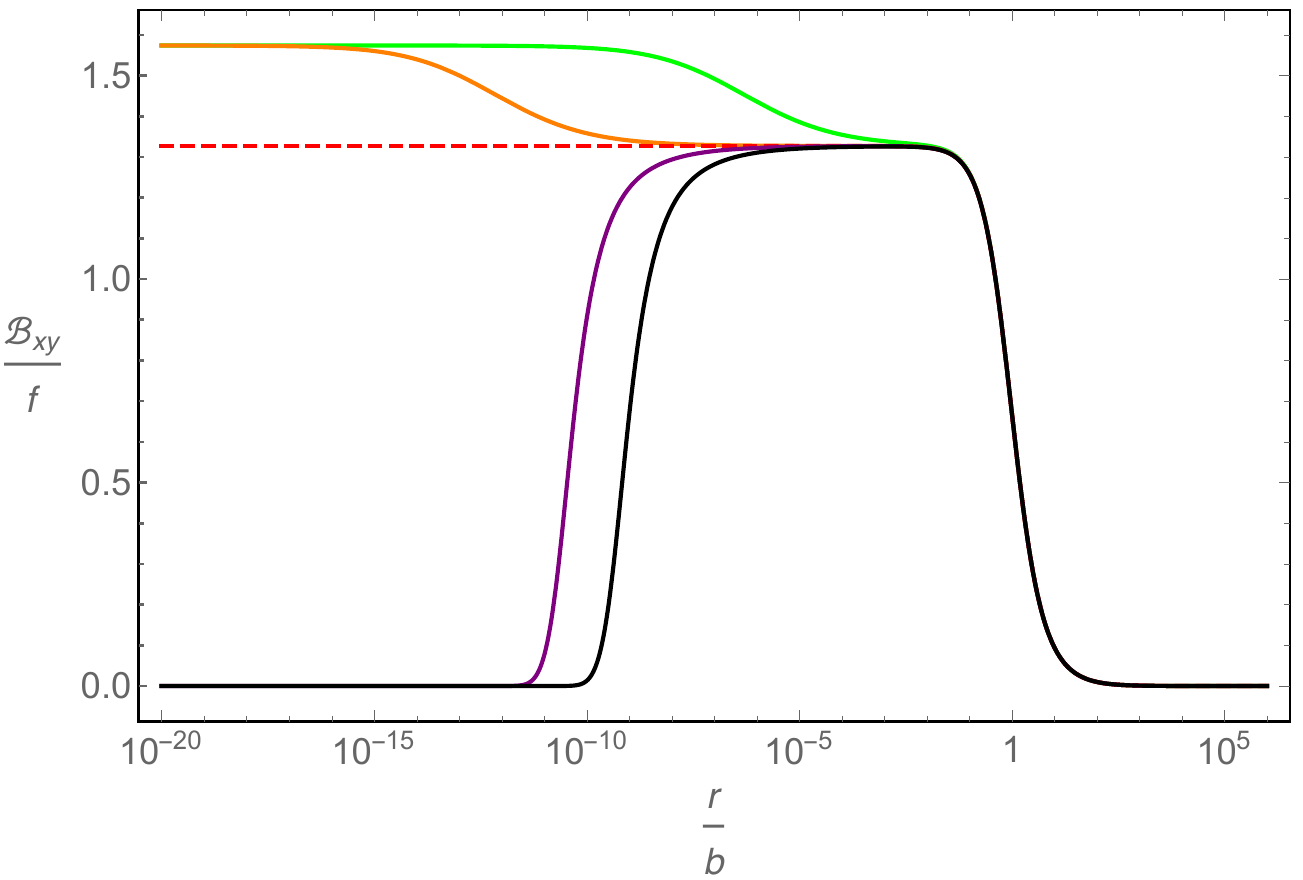}
  \end{minipage}
  \caption{\small Bulk profiles for background fields as a function of the radial coordinate in three different phases in the vicinity of critical point. Different colors are for different values of $\hat{M}=M/b$.}
  \label{fig:background}
\end{figure}

The profiles of $\phi,\, \mathcal{B}_{tz}/u$  and $\mathcal{B}_{xy}/f$ for parameters close to $(M/b)_c$ as a function of radial coordinate are illustrated  in Fig. \ref{fig:background}. 
As we gradually tune $M/b$ these profiles change smoothly from boundary to an intermediate scale $r/b \propto 10^{-5}$. 
However, the matter fields flow discontinuously to different types of IR profiles. 
Close to $(M/b)_c$, the geometry for three phases first flows to an intermediate Lifshitz type geometry from UV and then splits into different types of IR geometry. 
\vspace{.3cm}
 
We can compare the background geometries in the holographic models with and without the imposed the self-dual constraint. 
There is an extra dynamical component in $B_{ab}$  
in the improved NLSM model. The profiles of  $\phi$ and $B_{xy}$, as well as the type of the emergent symmetries at low-energy are similar to the model \cite{Liu:2018bye} in all the three phases. 
Then we come to the question what 
the topological properties are in this improved NLSM model, 
which will be studied in the next sections where we studied the properties of fermionic operator by probing massive fermions in the bulk.

\begin{figure}[h!]
  \centering
  \includegraphics[width=0.45\textwidth]{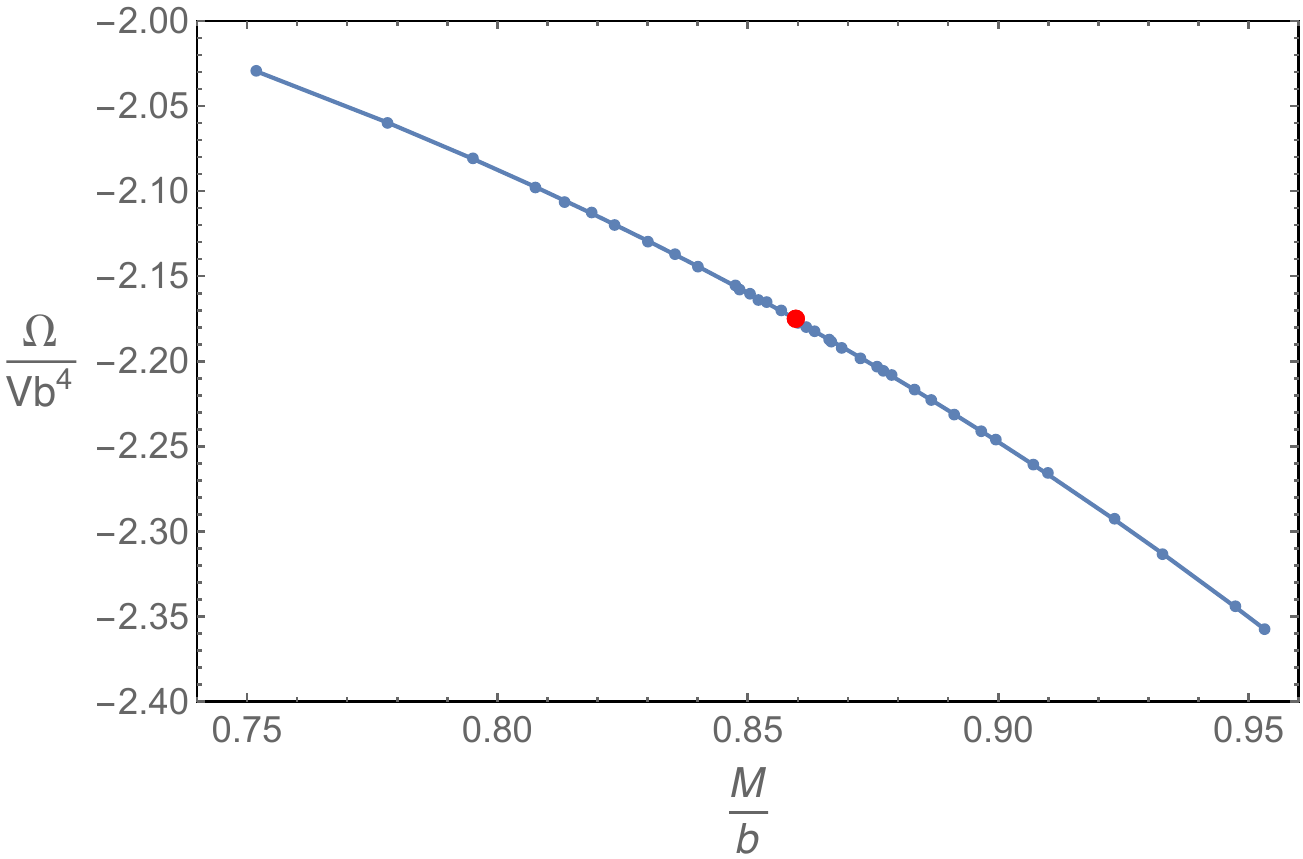}
  \caption{\small The free energy density as a function of 
  $M/b$ across the critical point (the red dot). The free energy density is continuous and smooth during the topological quantum phase transition.}
  \label{fig:freeE}
\end{figure}

Finally, the dependence of the free energy density as a function of $M/b$ in this system is shown in Fig. \ref{fig:freeE}. We find that the free energy density smoothly across the quantum phase transition point as $M/b$ increase and the first derivative of the free energy density with respect to $M/b$ reaches the same value in the vicinity of $(M/b)_c$ from two phases. Therefore, we conclude the topological phase transition in this improved NLSM is a continuous phase transition, while the self-duality does not change the order of the transition.

\section{ARPES from probe fermion}
\label{sec:arpes}
The remaining of this paper is to provide evidence of the topological band structure in the improved holographic nodal line semimetal and the existence of quantum phase transition. 
In experiments, angle resolved photoemission spectroscopy (ARPES) has been used to discover nodal band structure. 
Theoretically, this motivates us to investigate the low-energy effective topological Hamiltonian in the holographic nodal line semimetal phase, which is defined from the zero frequency Green's function of the Dirac fermionic operators, i.e., $H_t(\vec{k})=-G^{-1}(0, \vec{k})$ \cite{Wang:2012sm, Wang:2012ig}. From the topological Hamiltonian one could obtain the topological properties of the nodal lines.
 
In the following we shall use the same strategy as 
\cite{Liu:2018bye} to study the fermionic spectral function by probing a fermion in the bulk of our model. We shall compare the results of this improved model and the previous model without duality condition of the two form operators. 

\subsection{Holographic fermionic spectral function}
In holography, we can obtain the fermionic spectral function by probing a single fermion in the 
gravitational bulk geometry \cite{Cubrovic:2009ye, Liu:2009dm}. 
However, in the four dimensional boundary field theory with a five dimensional bulk dual, the Dirac fermion in the bulk corresponds to chiral fermionic operator in the field theory 
\cite{Iqbal:2009fd, Liu:2018bye, Plantz:2018tqf}. 
To obtain a Dirac fermionic operators 
in the boundary field theory with nontrivial spectral structure, in bulk we need to consider two sets of Dirac fermions coupled with each other through the scalar field $\Phi$ and the self-dual two form field $B_{ab}$ \cite{Liu:2018bye} with the following action
\bea
\label{eq:bulkdirac5d}
\begin{split}
S_{\text{fermion}}&=S_1+S_2+S_{\text{int}}+S_{\text{bdy}}\,,\\
S_{\text{int}}&=S_{\Phi}+S_{B}\,,
\end{split}
\eea
where $S_{\text{bdy}}$ is the boundary term to make the theory self-consistent and 
\bea
\begin{split}
S_1&=\int d^5x\sqrt{-g}~\bar{\Psi}_1\left(\Gamma^aD_a-m_f  \right)\Psi_1\,,\\
S_2&=\int d^5x\sqrt{-g}~\bar{\Psi}_2\left(\Gamma^aD_a+m_f  \right)\Psi_2\,
\end{split}
\eea
are the action for two types of free fermions with opposite signs of mass and with different quantizations where $\bar{\Psi}=\Psi^\dagger i \Gamma^0$ and $D_a=\nabla_a-iq_3A_a$,  
while
\bea
\begin{split}
S_{\Phi}&=-\int d^5x\sqrt{-g} \left( \eta_1\Phi\bar{\Psi}_1\Psi_2 +  \eta^*_1\Phi^*\bar{\Psi}_2\Psi_1 \right)\,,\\
S_{B}&=\int d^5x \sqrt{-g}\left(\eta_2 B_{ab} \bar{\Psi}_1\Gamma^{ab}\gamma^5\Psi_2-\eta^*_2B^*_{ab}\bar{\Psi}_2\Gamma^{ab}\gamma^5\Psi_1\right)\,
\label{eq:fermioninteraction}
\end{split}
\eea
describe how the two types of fermions couple to the background scalar field $\Phi$ and the two form field $B_{ab}$, respectively. The details of the boundary term and the definitions of gamma matrices in the bulk are shown in the appendix \ref{appD}. 
In the absence of the interaction terms, there are two free Dirac fermions in the bulk leading to two independent sets of chiral fermions in the field theory and therefore the interactions are necessary. 
These interaction terms take similar forms as that in the field theoretical model, although there are several other possible interaction terms. For example, one may also couple the two fermions to $B_{ab}$ with no insertion of $\gamma^5$. 
The reason for this choice is to produce the nodal band structure and corresponding topological properties of the nodal line semimetal \cite{Liu:2018bye}.

The equations of motion for these two fermions are
\bea
\label{eq:diracfermions}
\begin{split}
\left(\Gamma^aD_a-m_f\right)\Psi_1-\eta_1\Phi\Psi_2+\eta_2B_{ab}\Gamma^{ab}\gamma^5\Psi_2&=0\,,\\
\left(\Gamma^aD_a+m_f\right)\Psi_2-\eta_1^{*}\Phi^{*}\Psi_1-\eta_2^{*}B^{*}_{ab}\Gamma^{ab}\gamma^5\Psi_1&=0\,.
\end{split}
\eea
After taking the ansatz
\bea
\begin{split}
\Psi_l&=(-gg^{rr})^{-1/4}\psi_le^{-i\omega t+ik_xx+ik_yy+ik_zz}
=(uf)^{-1/2}\psi_le^{-i\omega t+ik_xx+ik_yy+ik_zz}\,,
\end{split}
\eea
where $l=1,2$ and $\psi_l=\psi_l(r)$ are only functions of radial coordinate, 
(\ref{eq:diracfermions}) can be simplified as 
\bea
\begin{split}
\label{eq:fermion-eom}
\left(\Gamma^{\underline{r}}\partial_r+\frac{1}{u}\left(-i \omega \Gamma^{\underline{t}}+i k_z\Gamma^{\underline{z}} \right) +\frac{1}{\sqrt{uf}} \left(i k_x\Gamma^{\underline{x}} + i k_y\Gamma^{\underline{y}}\right) 
+(-1)^l\frac{m_f}{\sqrt{u}}\right)\psi_l ~~~&\,\\
-\left(\eta_1\frac{\phi}{\sqrt{u}}+2\eta_2\left((-1)^l\frac{\mathcal{B}_{xy}}{\sqrt{u}f}\Gamma^{\underline{xy}}  - i
\frac{\mathcal{B}_{tz}}{\sqrt{u}u}\Gamma^{\underline{tz}} \right)\gamma^5\right)\psi_{3-l}&=0\,.
\end{split}
\eea
In this equation, the couplings in the second line mixed the two different fermions and we have assumed that the couplings $\eta_1$ and $\eta_2$ are real numbers. 
Note that comparing with \cite{Liu:2018bye}, the last term in (\ref{eq:fermion-eom}) is new. 
However, this new term does not change the asymptotic behaviors of this equation both in the IR and UV limits. 

In the following, we introduce the 
necessary steps in calculating the fermionic spectral function including the boundary conditions 
in both IR and UV regime \cite{Liu:2018bye, Iqbal:2009fd}. 
We also define the topological Hamiltonian and its related eigenstates in terms of the fermionic spectral function from holography \cite{Liu:2018bye}. 
To keep the main text straightforward, we take the NLSM phase as the example and leave details in other phases in appendix \ref{appD}.

The system (\ref{eq:fermion-eom}) contains eight first order, non-homogeneous differential equations and we shall solve them numerically with proper boundary conditions. 
We start from the asymptotic IR region $(r\rightarrow 0)$, where the nonlinear (coupling) terms become irrelevant and the equations can be analytically solved to leading order. 
For example in the nodal line semimetal phase with $\omega$ or $k_z$ finite, when $r\rightarrow 0$, 
\be
\frac{1}{u}\propto \frac{1}{r^2}\,,~~~~ \frac{1}{\sqrt{uf}}\propto \frac{1}{r^{1+\alpha/2}}\,,~~~~ \frac{\mathcal{B}_{xy}}{\sqrt{u}f}\propto \frac{\mathcal{B}_{tz}}{\sqrt{u}u}\propto \frac{1}{r}\,.
\ee
Since $\psi_1, \psi_2$ are in the same order of $r$, then the equations can be simplified into decoupled forms describing two independent free fermions
\bea
\label{eq:fermionir}
\left(\Gamma^{\underline{r}}\partial_r + \frac{1}{u_0 r^2}\left(-i\omega\Gamma^{\underline{t}}+i k_z \Gamma^{\underline{z}}  \right)  \right)\psi_l=0\,. 
\eea
Using the 2-components spinors $\psi_{l\pm}$ defined in appendix \ref{appD}, 
the first-order spinor differential equations can be 
simplified to second order scalar differential equations
\bea
\frac{u_0^2r^4}{\omega^2-k_z^2}\partial_r^2\psi_{l+}+\frac{2u_0^2r^3}{\omega^2-k_z^2}\partial_r\psi_{l+}+\psi_{l+}=0\,,
\eea
while $\psi_{l-}$ is determined by $\psi_{l+}$. 
After taking the explicit representation of the gamma matrices introduced in the appendix \ref{appD}, one can obtain the leading order infalling solutions analytically expressed as 
\bea
\label{eq:IR fermion}
\psi_{l}^\text{IR}\simeq e^{i\frac{\sqrt{\omega^2-k_z^2}}{u_0r}}
\left(
  \begin{array}{c}   
   z_1^l \,(1+...)\\ 
   z_2^l \,(1+...)\\
   i\frac{\sqrt{\omega^2-k_z^2}}{\omega-k_z}\,z_1^l\,(1+...)\\
   i\frac{\sqrt{\omega^2-k_z^2}}{ \omega+k_z}\,z_2^l \,(1+...)\\
  \end{array}
\right)
\eea
where $z^l_1, \,z^l_2$ are free constants, and the dots represent the higher order corrections with respect to $r$. 
Note that these IR behavior is the same as the one in \cite{Liu:2018bye} and this is due to that the scaling symmetry of the IR geometry in NLSM  here is the same as 
\cite{Liu:2018bye}. 

Close to the UV boundary, similar to the case in the IR region, several terms become irrelevant and the equations take  
a simple form 
\bea
\begin{split}
\left(r\Gamma^{\underline{r}}\partial_r-m_f\right)\psi_1=0\,,
~~~~
\left(r\Gamma^{\underline{r}}\partial_r+m_f\right)\psi_2=0\,.
\end{split}
\eea
The masses determine the conformal dimensions of the fermionic operators, while our choice of coupling forms $S_\Phi$ and $S_B$ do not change the conformal dimensions. In components 
the Dirac fields can be solved as
\bea
\label{eq:UV fermion}
\psi_{1}=
\left(
  \begin{array}{c}   
   s_1~ r^{m_f}+...\\ 
   s_2~ r^{m_f}+...\\
   r_3 ~r^{-m_f}+...\\
   r_4 ~r^{-m_f}+...\\
  \end{array}
\right)\,,
~~~~
\psi_{2}=
\left(
  \begin{array}{c}   
   r_1~ r^{-m_f}+...\\ 
   r_2~ r^{-m_f}+...\\
   s_3~ r^{m_f}+...\\
   s_4~ r^{m_f}+...\\
  \end{array}
\right)
\eea
where the coefficients $s_i$ and $r_i$ 
depend on the choices of $z^l_1$ and $z^l_2$ and can be solved numerically.

In this paper we will focus on the case with $m_f=-\frac{1}{4}$. 
\footnote{We expect our results are independent of the choice of $m_f$ since the near horizon boundary condition does not depend on the mass parameter. Nevertheless, it is still interesting to perform explicit numerical study to investigate whether richer effects might appear if we vary $m_f$.} 
We take alternative quantization for $\psi_1$ while  standard quantization for $\psi_2$, i.e., in both cases the dominant modes that proportional to $r^{-m_f}$ as the operator while the subdominant modes proportional to $r^{m_f}$ as the external source. 
The source and the response are
\bea
\label{eq: sourceresponse}
\psi_s
=
\left(
  \begin{array}{c}   
   s_1\\ 
   s_2\\
   s_3\\
   s_4\\
  \end{array}
\right)\,,~~~~~~
\psi_r
=
\left(
  \begin{array}{c}   
   -r_1\\ 
   -r_2\\
   r_3\\
   r_4\\
  \end{array}
\right)\,.
\eea
With the infalling boundary conditions (\ref{eq:IR fermion}) in IR, (\ref{eq:fermion-eom}) can be solved numerically and we can get $\psi_s$ and $\psi_r$.  
The spinor operator $\psi_r$ and the spinor source $\psi_s$ 
are related by 
$
\psi_r(k)=-i\Xi(k) \psi_s(k)
$ 
where $\Xi(k)$ is a $4\times4$ matrix. 
To compute the matrix $\Xi(k)$, we need at least four sets of linearly independent 
sources and repsonses
which can be obtained 
with four independent infalling boundary conditions at IR. 
We have a matrix equation for $\Xi(k)$
\bea
\label{reqXs}
M_r=-i\Xi M_s\,,
\eea
 where
\bea
M_r=\begin{pmatrix}
  \psi^{\uppercase\expandafter{\romannumeral1}}_r, & \psi^{\uppercase\expandafter{\romannumeral2}}_r, & \psi^{\uppercase\expandafter{\romannumeral3}}_r, & \psi^{\uppercase\expandafter{\romannumeral4}}_r\\ 
\end{pmatrix}\,,~~~
M_s=\begin{pmatrix}
  \psi^{\uppercase\expandafter{\romannumeral1}}_s, & \psi^{\uppercase\expandafter{\romannumeral2}}_s, & \psi^{\uppercase\expandafter{\romannumeral3}}_s, & \psi^{\uppercase\expandafter{\romannumeral4}}_s\\ 
\end{pmatrix}\,
\eea
are $4\times 4$ matrices. 
By right multiplying $M_s^{-1}$ on both sides of (\ref{reqXs}), one obtains $\Xi=iM_r M_s^{-1}$, leading to the retarded Green's function for the fermionic operator
\bea
G_R(k)=i\Gamma^{\underline{t}}M_r(k) M_s^{-1}(k)\,.
\eea

The effective Hamiltonian is defined as 
\bea
\label{Eq:Heff}
H_{\text{eff}}(\vec{k})\equiv -G^{-1}_R(0,~\vec{k})\,,
\eea
which is a generalization of topological Hamiltonian that is first introduced as a probe to detect topological invariants in topological insulators \cite{Wang:2012sm, Wang:2012ig}. We will study the topological invariants from this topological Hamiltonian in section \ref{sec:ti}. 

\subsection{Numerical results}
\label{subsec:nr}

In the previous subsection, we have introduced 
how to calculate the fermionic spectral function from holography. 
In the following, we 
explain our numerical results. 

Similar to the case without self-duality in \cite{Liu:2018bye},  we find that there exist multiple Fermi surfaces in the holographic nodal line semimetal phase and the dispersion close to the Fermi surface is linear. This observation is obtained from numerical calculation of the eigenvalues of the effective Hamiltonian (\ref{Eq:Heff}) from which the band structure at $k_z=0$ plane in the momentum space is studied.

\begin{figure}[h!]
  \centering
  \includegraphics[width=0.600\textwidth]{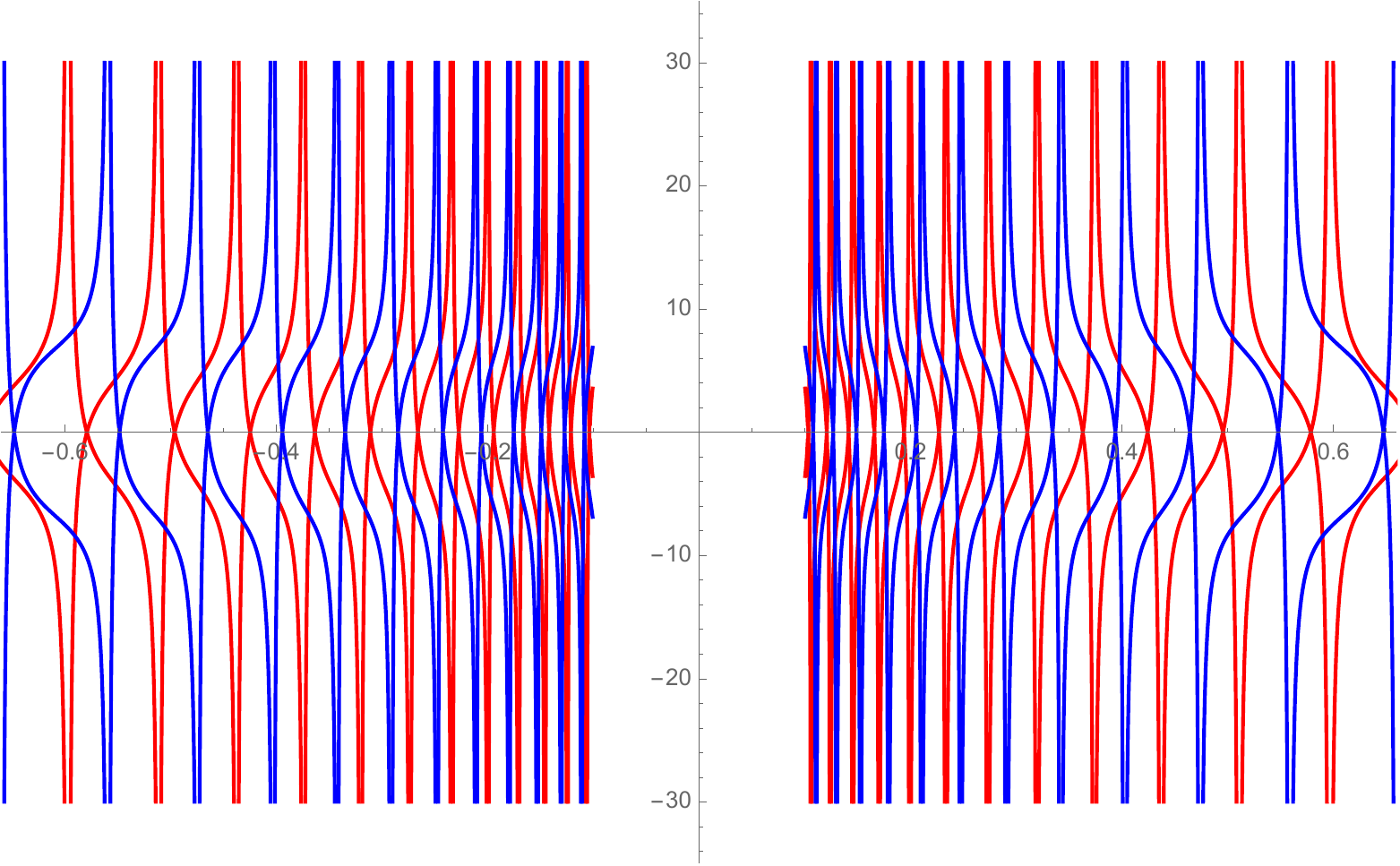}
  \caption{\small Eigenvalues of $G^{-1}(0,k_x,k_y=k_z=0)$ in the holographic NLSM phase as a function of $k_x$ for $M/b \simeq 0.0014$, in which blue and red curves represent two different sets of eigenvalues separately. 
Therefore one concludes that multiple Fermi surfaces exist in the NLSM phase. }
  \label{Band}
\end{figure}

More precisely, since we have assumed the existence of $SO(2)$-symmetry in the $x$-$y$ plane, we fix $k_y=0$ and compute $H_{\text{eff}}(k_x, k_y=k_z=0)$ without loss of generality and the locations of $H_{\text{eff}}=0$ indicates the nodal circle of Fermi surfaces. 
The four eigenvalues appear in two pairs and can be arranged as $\{h_1, -h_1, h_2, -h_2\}$ with $h_1, h_2\geq 0$ varies as a function of $k_x$. 
We show the effective band structure in the nodal line semimetal phase for $M/b\simeq 0.0014$ in Fig. \ref{Band}, where the red and blue curves are used to represent two different groups of bands. 
For convenience, we use ``Band-1" and ``Band-2" to describe the bands in red and blue curves in the following. 
The most interesting observation from this effective band structure is the existence of multiple and discrete Fermi surfaces\footnote{Multiple Fermi surfaces in holography has also been found in \cite{Hartnoll:2011dm, Iqbal:2011in, Cubrovic:2011xm} for a finite density system.} for both energy bands indicated by band crossing at $h_1=h_2=0$. 
These Fermi surfaces appear alternately in ``Band-1" and ``Band-2", and more densely as the momentum $k_x$ decrease. 
We have checked that this characteristic band structure generically exist in the nodal line semimetal phase, not limited to particular values of $M/b$. Then a natural question arise that, are these band crossings accidental or topologically nontrivial? To answer this question, one should study the topological invariant and 
we leave to the next section. 

The existence of nodal line shaped Fermi surface is the signature of NLSM. In the weakly coupled model, the Fermi surface appears at $k_z=0$, while for finite $k_z$ there exists only gapped band structure. 
Fig. \ref{Band} shows that there exist multiple Fermi surfaces at $k_z=0$ in the strongly-coupled NLSM.  Next  we will explain that there is no Fermi surface for finite $k_z$ from the analysis of the fermionic spectral function. 
Note that the fermionic spectral function is defined from the imaginary part of the retarded Green's function $G_R(\omega, \vec{k})$, from which a Fermi surface can be identified when there is a sharp peak in $\text{Im}[G_R(\omega\rightarrow 0, \vec{k})]$ \cite{Cubrovic:2009ye, Liu:2009dm}. 
The IR boundary condition for the fermions in the NLSM phase is shown is appendix \ref{appD4} and one concludes  that 
\begin{itemize}
 \item for $k_z=0$ while $\omega\rightarrow 0$,  the IR boundary condition is complex, which leads to the complex $G_R(\omega\rightarrow 0, \vec{k})$\,;
 \item for finite $k_z$ while $\omega\rightarrow 0$, the IR boundary condition is pure real and there is no imaginary part in $G_R(\omega\rightarrow 0, \vec{k})$\,. 
\end{itemize}
Therefore, there is no Fermi surface when $k_z\neq 0$ since the spectral function vanishes. In addition, it is a special case when $\omega=0$ and we can identify the Fermi surface via the effective Hamiltonian \eqref{Eq:Heff} that gives consistent results as that from the spectral function.

\begin{figure}[h!]
  \centering
  \includegraphics[width=0.56\textwidth]{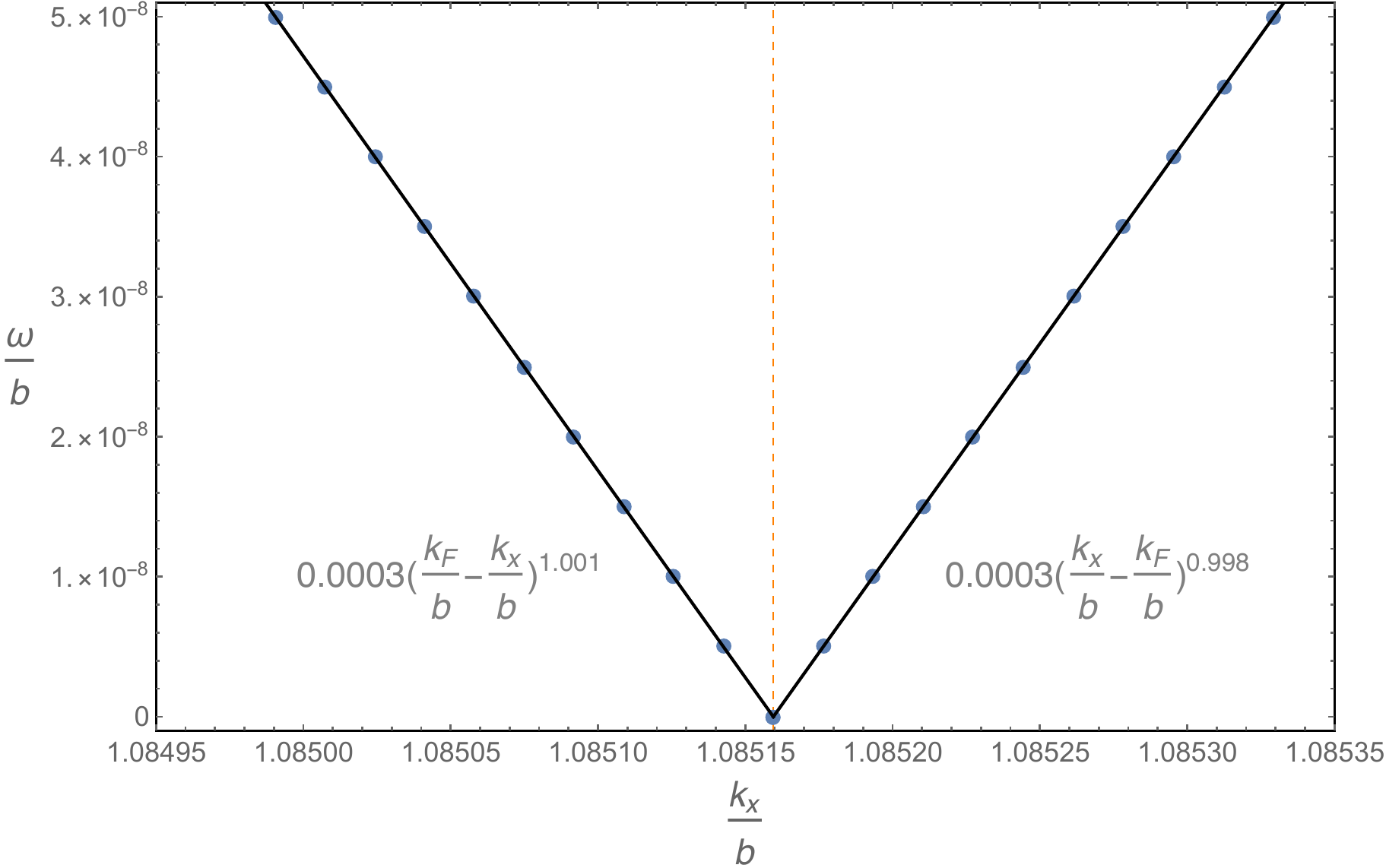}
  \caption{\small The linear dispersion around one of the multiple Fermi surfaces at $k_F/b=1.085$ for $\frac{M}{b}\simeq 0.0014$. 
  }
  \label{fig:dispersion}
\end{figure}

In the vicinity of the Fermi surface, the dispersion relation of the excitations is linear in $k_x$. 
We make a generalization for the effective Hamiltonian  (\ref{Eq:Heff}) to 
finite $\omega\ll 1$ case. From the locations of zeros in $H_{\text{eff}}$  the dispersion relation can be obtained. 
As shown in Fig. \ref{fig:dispersion}, we choose several discrete frequencies, get the locations of zeros and then plot these points in the dimensionless $\omega$-$k_x$ plane. 
We use power law functions to fit the data and find that close to the Fermi surface in the small $\omega$ region, the almost linear function perfectly fit the data. The linear dispersion property of the excitations around a circular Fermi surface is another evidence to indicate that the ground state is a nodal line semimetal.
\begin{figure}[h!]
  \centering
  \includegraphics[width=0.56\textwidth]{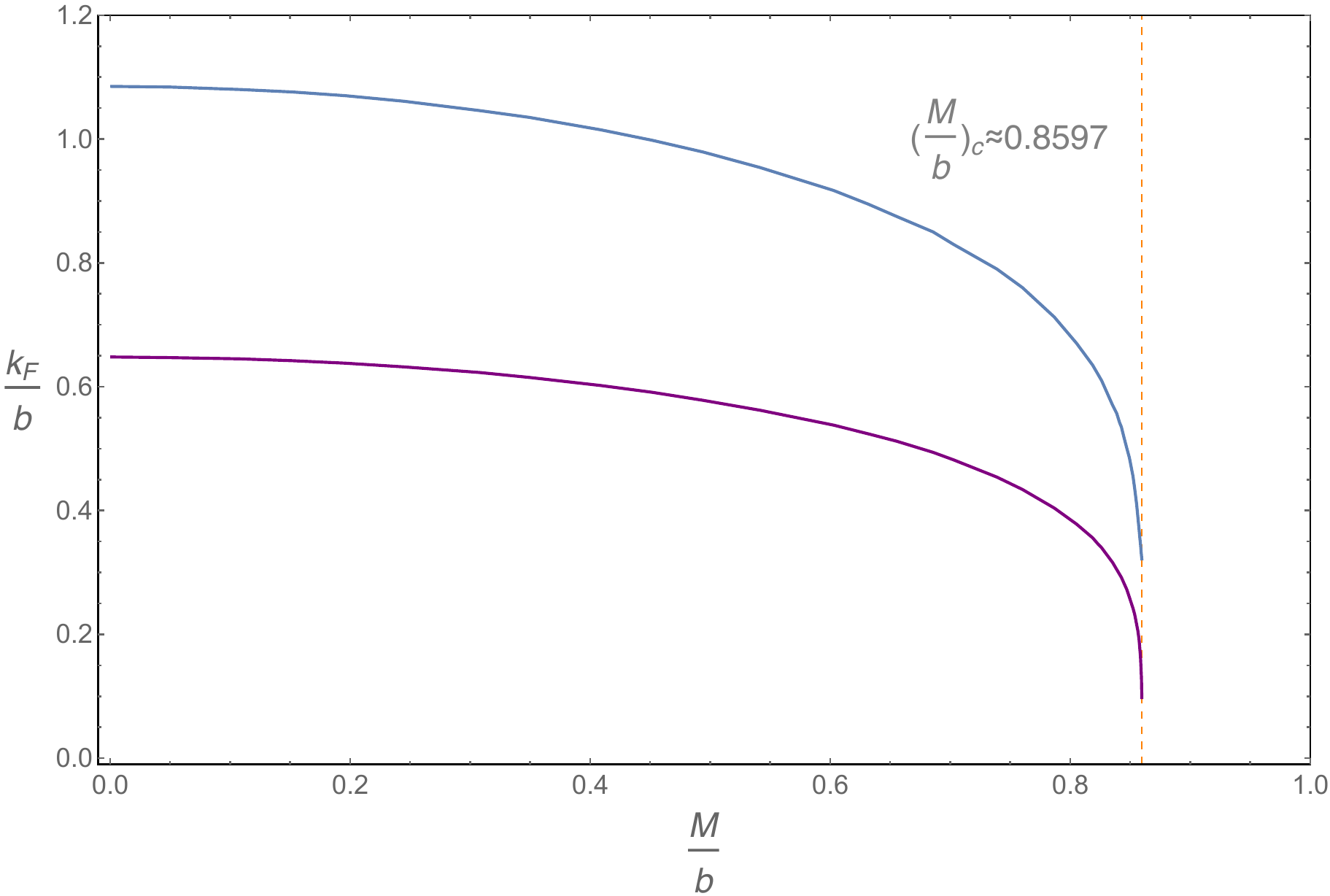}
  \caption{\small The size of fermi surfaces shrinks as $M/b$ increases. Fermi surfaces can exist in the nodal line semimetal phase as well as at the critical point while disappear in the trivial phase. The sudden disappearance of Fermi surfaces can be viewed as the consequence of the holographic quantum phase transition. }
  \label{fig:FSshrink}
\end{figure}

The quantum phase transition from the topological nodal line semimetal phase to a trivial phase can also be reflected from the effective band structure. 
This can be observed that  
the size of nodal circles $k_F=\sqrt{k_x^2+k_y^2}$ shrink as $M/b$ increases. {\footnote{An interesting observation is that, the size of Fermi surfaces shrink in the holographic superfluids when the temperature is increased, e.g. in \cite{Gubser:2010dm, Ammon:2010pg}. In the zero temperature topological NLSM phase, the size of Fermi surfaces shrinks as $M/b$ changes instead of temperature. }
We demonstrate the shrinking of two different Fermi surfaces in Fig. \ref{fig:FSshrink} where $k_F/b$ is a smooth, monotonic decreasing function when $0<M/b<0.8597$, which is a generic feature for all the Fermi surfaces and indicates that multiple Fermi surfaces cannot be removed from small perturbations of the ground state. 
In the topological trivial phase, there is no Fermi surface other than $k_F=0$ since from the near horizon we know that  the retarded fermionic Green's function is real for all values of spacelike $k^\mu$ \cite{Iqbal:2009fd, Liu:2018bye}. 
Finally, $k_F$ does not vanish at the critical point, which is different from the weakly-coupled theory. 
The reason is that, even though the form of interactions in  (\ref{eq:fermioninteraction}) is a general approach to realize the nodal circle band structure, the couplings also deform the shape of the spectra. 
Therefore, it is reasonably expected that by fine-tuning $\eta_1$ and $\eta_2$, the Fermi surface could exactly shrink to zero at the critical point. 

\section{Topological invariant in nodal line semimetal}
\label{sec:ti}
In the previous section, we have studied the effective band structure of the strongly coupled nodal line semimetal from the topological Hamiltonian of 
the fermionic Green's function in holography. 
The results, especially the existence of multiple nodal lines and linear dispersion close to each nodal line, show many similarities to the case without self-duality \cite{Liu:2018bye}. 
In this section we will 
further study the topological property of these nodal lines. 

We will study the  
topological invariants in the holographic semimetal phase to tell whether the  
nodal lines can be removed by small perturbations. In the nodal line semimetal system, one of the topological invariants is Berry phase which is defined on a closed path enclosing a 
node along the line \cite{rev1, Berry:1984}. 
By ``enclosing" we mean that the path do not touch the node and the node cannot get out of the closed loop without cutting down the loop. 
The nodal line is accidental when the Berry phase associated with a closed loop enclosing 
a node on this line 
is 0 while topological protected if the Berry phase is $\pi$. 
In the following, we will discuss the Berry phase in both field theory and holography. 

\subsection{Topological invariant  in field theory}

We first calculate Berry phase from the weakly coupled field theory where the eigenvalues and eigenstates can be obtained analytically. 
Without loss of generality we set 
$b_{xy}=1/2, m=0$ in the Lagrangian (\ref{eqL:imNLSM}). With these parameters the system is in the nodal line semimetal phase and the nodal line is located on the $k_x$-$k_y$ plane with $k_z=0$ in the momentum space. 
The choice of the closed path along which we shall compute the Berrry phase is shown in Fig. \ref{fig:enclosedloop}. 
Note that we have the nodal line which is described by the red circle, and each point on this circle is a Weyl node. We focus on one single Weyl node and use a closed loop to enclose this point. 
For example, we choose $k_y=0$ and the closed path can be parametrized as 
\be
\label{eq:discrete}
\left(k_x,~k_z\right)=\left(k_F+l~\text{sin}~\theta,~l~\text{cos}~\theta\right)\,,~~~~\text{with} ~~\theta \in\left[0,~2\pi\right)\,.
\ee 
Since we will generalize the discussion to strongly coupled theory in which the eigenstates can only be obtained at discrete points, we select a series of discrete points along the path, which is illustrated in the right plot in Fig. \ref{fig:enclosedloop}. We have $\theta$ in (\ref{eq:discrete}) with $\theta_i=\frac{2\pi i}{N}$ with $i\in \{1,..., N\}$. 
\vspace{.3cm}
\begin{figure}[h!]
  \centering
\includegraphics[width=0.5\textwidth]{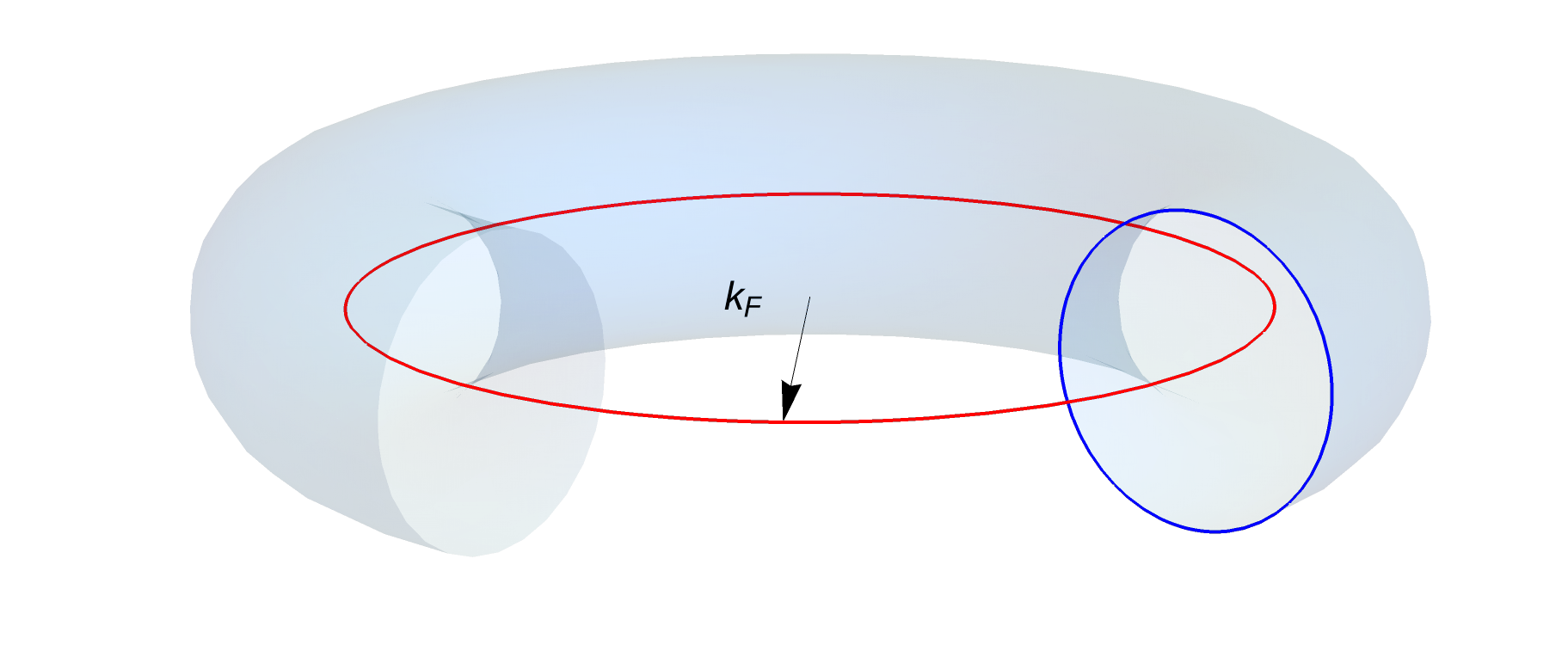}
\includegraphics[width=0.22\textwidth]{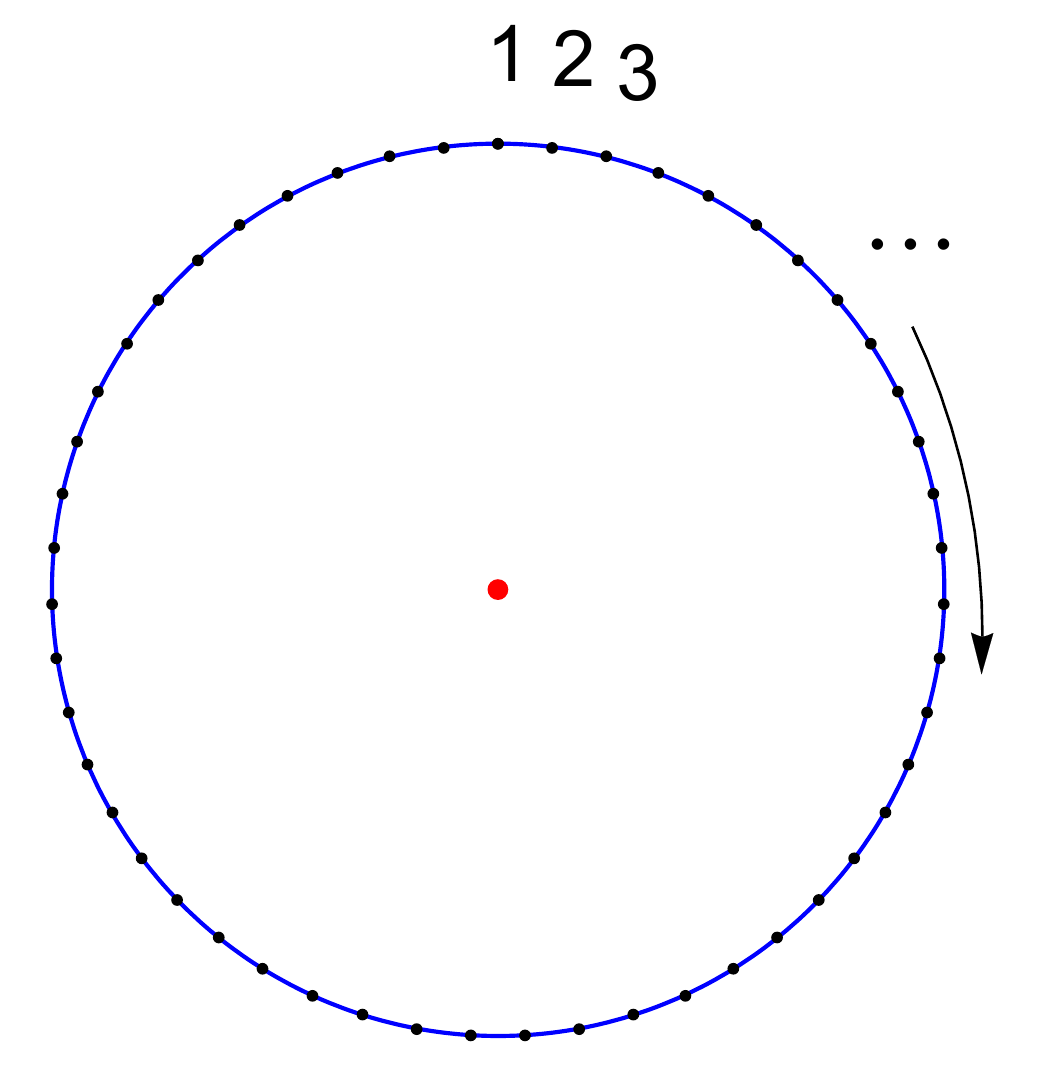}
  \caption{\small Left: The illustration of the closed path (blue curve) that encloses a Weyl node along the nodal line (red curve). Right: A series of discrete black points are selected along the closed path.} 
  \label{fig:enclosedloop}
\end{figure}

By making the circle discrete as above, we can define the discrete Berry phase
\be
\label{eq:bpdis}
e^{-i\phi_{i_1i_2}}=\frac{\langle n_{i_1}|n_{i_2}\rangle}{|\langle n_{i_1}|n_{i_2}\rangle|}\,,
\ee
where $| n_{i_1}\rangle$ and $| n_{i_2}\rangle$ are eigenstates at two adjacent discrete points $i_1$ and $i_2$ along the path. 
The total Berry phase along the closed path is then the summation of all the discrete phases. 

When 
$k_y=0$ from the Lagrangian (\ref{eqL:imNLSM}) and the parameters we chose above, 
we have the eigenstates
\bea
\label{eq:esQFT}
\begin{split}
|n^I\rangle&=\left(1\,,~~\frac{k_z+\sqrt{(k_x-2)^2+k_z^2}}{2-k_x}\,,~~\frac{k_z+\sqrt{(k_x-2)^2+k_z^2}}{k_x-2}\,,~~1\right)^T\,,\\
|n^{II}\rangle&=\left(1\,,~~\frac{k_z-\sqrt{(k_x-2)^2+k_z^2}}{2-k_x}\,,~~\frac{k_z-\sqrt{(k_x-2)^2+k_z^2}}{k_x-2}\,,~~1\right)^T\,,
\end{split}
\eea
and the associated energy eigenvalues $E^I=-\sqrt{(k_x-2)^2+k_z^2}$,~ $E^{II}=\sqrt{(k_x-2)^2+k_z^2}$. 
For the choice of the ``discrete" circle with (\ref{eq:discrete}), we have either states with energy $-l$ or $l$. 
By substituting (\ref{eq:discrete}) and (\ref{eq:esQFT}) into (\ref{eq:bpdis}) and do the summation of all the discrete phases, we find that the nontrivial phase factor is from the point where the normalized norm of two adjacent eigenstates becomes $-1$. For states with energy $-l$, i.e. $|n^I\rangle$,  due to $\langle n^I(k_x=2+0_-,k_z\simeq l)|n^I(k_x=2+0_+,k_z\simeq l)\rangle <0$ we have a phase factor $\pi$ from (\ref{eq:bpdis}). Similarly for states with energy $l$, i.e. $|n^{II}\rangle$, the phase factor $\pi$ of the Berry phase is due to $\langle n^{II}(k_x=2+0_-,k_z\simeq -l)|n^{II}(k_x=2+0_+,k_z\simeq -l)\rangle <0$. Therefore one concludes that Berry phase associated with a closed loop enclosing the node is $\pi$, which means that the nodal line semimetal described by (\ref{eqL:imNLSM}) is topologically nontrivial.

\subsection{Topological invariant in holography}
In this subsection we compute Berry phase in the strongly coupled NLSM system from holography. 
Similar to the discussion in field theory, 
in holography we first compute the eigenstates of the effective Hamiltonian 
at a series of discrete points along a closed path in momentum space. The choice of discrete points and the definition of Berry phase are the same as the previous subsection for field theory, as shown in Fig. \ref{fig:enclosedloop}. 
Also, the Fermi surfaces only locate at the $k_z=0$ plane in the momentum space since the imaginary part of $G_R(\omega\rightarrow 0, \vec{k})$ vanishes for finite $k_z$, as explained in \ref{subsec:nr} in detail. 
The difference comparing to the field model is that, now there are multiple and dense nodal lines in the effective band structure. We should be careful that the closed loop encloses only a single Weyl node. 


Note that the IR boundary conditions become a little bit tricky in this case because $k_x$ and $k_z$ are finite simultaneously while the $SO(2)$ symmetry in the $k_x$-$k_z$ plane is broken at low energy. 
When $k_z$ is finite, for example in the same order compared to $k_x$, the $k_x$-term in (\ref{eq:fermion-eom}) can always be ignored since $r^{-1-\alpha/2}\ll r^{-2}$. Instead, for the discrete points close to the $k_x$-axis where $k_z$ is very small, in the deep IR   
the $k_x$-term in (\ref{eq:fermion-eom}) can not ignored any more and will modify 
the IR boundary conditions for the bulk fermions. 
However, whenever $k_z$ is nonzero one can always choose a sufficiently small $r$ such that the $k_z$ term dominates in (\ref{eq:fermion-eom}). Therefore one could expect that 
the modification from finite $k_x$ on the ingoing spinor wave will not change the Berry phase. We list the IR boundary conditions for the cases with nonzero $k_z$ and $k_z=0$ in appendix \ref{appD}. 
In the following we first show the numerical results of the eigenstates and eigenvalues along the ``discrete" circle. 

We numerically show the dependence of the value of four components in the eigenstates on momentum for the negative energy eigenvalues in Fig. \ref{fig:eigenstate}.
As shown in this figure, each component poses sudden jump 
when across the $k_x$-axis from the upper plane to lower or vice versa. 
This is a common numerical result which does not depend on the radius of the ``discrete" circle. Since the eigenstates change discontinuously across the $k_x$-axis, we should be careful to deal with points along the $k_x$-axis,  
i.e., to analyze the eigenstates for $k_z=0$ cases.  
\vspace{.3cm}
\begin{figure}[h!]
  \centering
\includegraphics[width=0.24\textwidth]{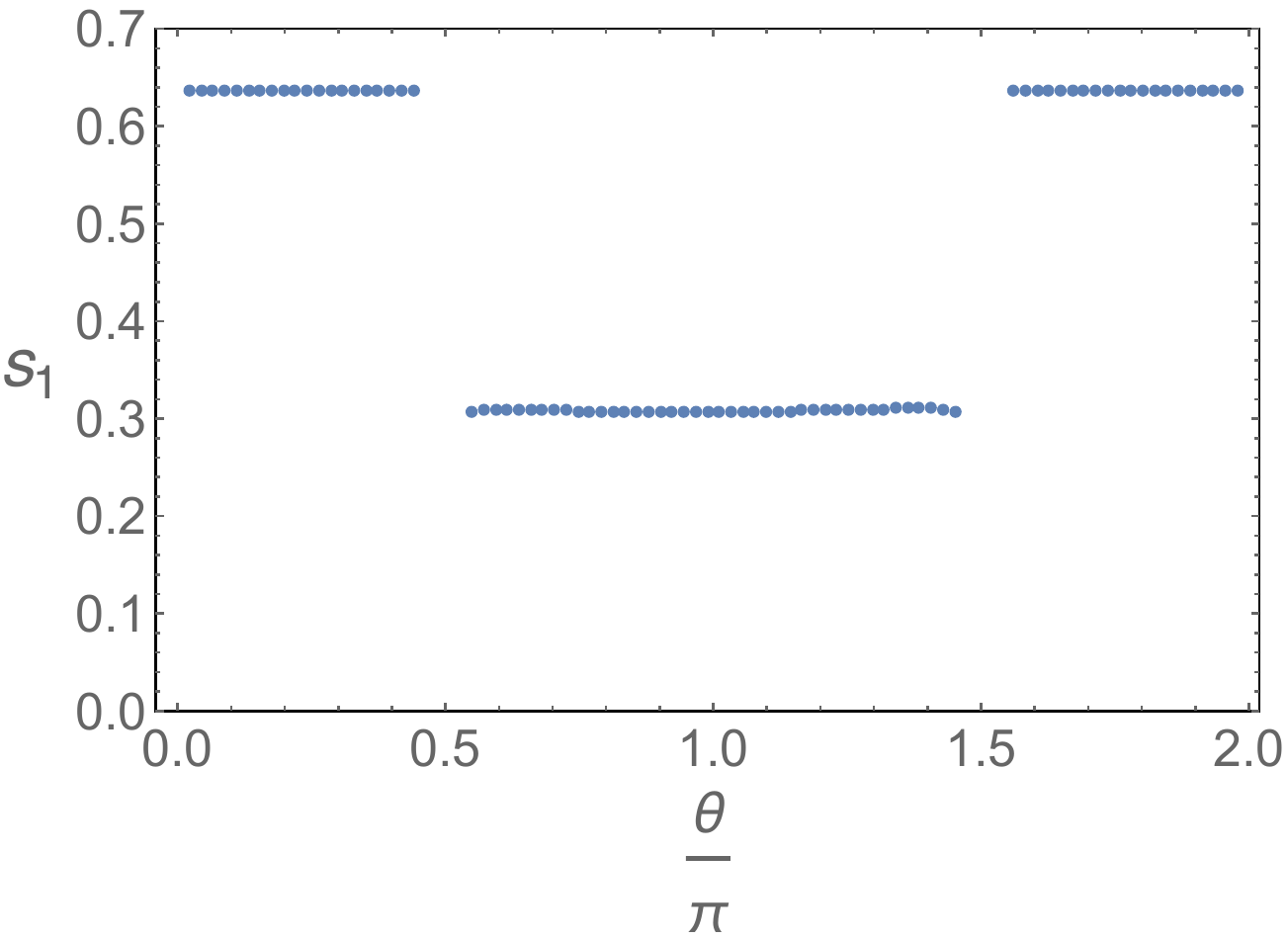}
\includegraphics[width=0.24\textwidth]{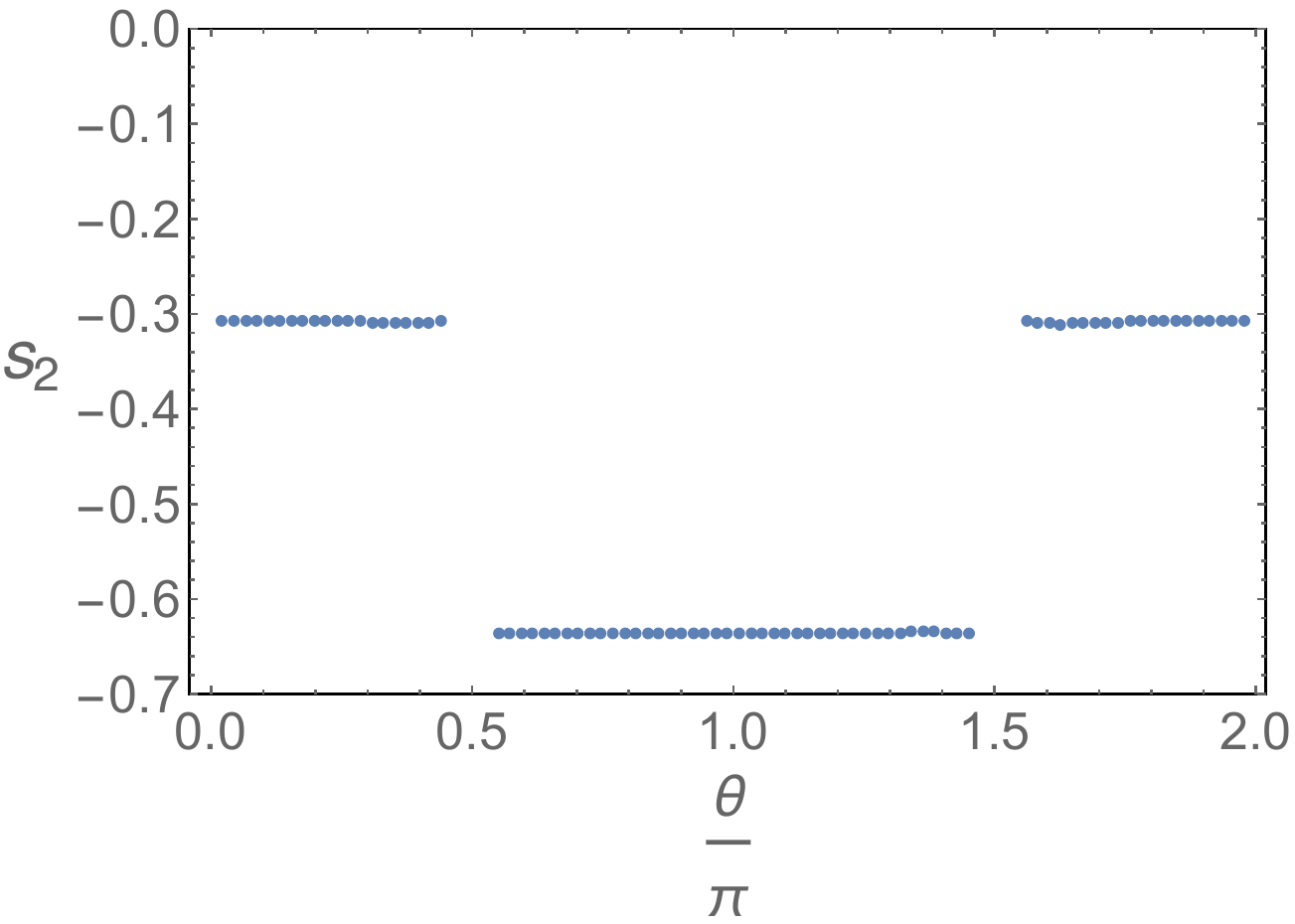}
\includegraphics[width=0.24\textwidth]{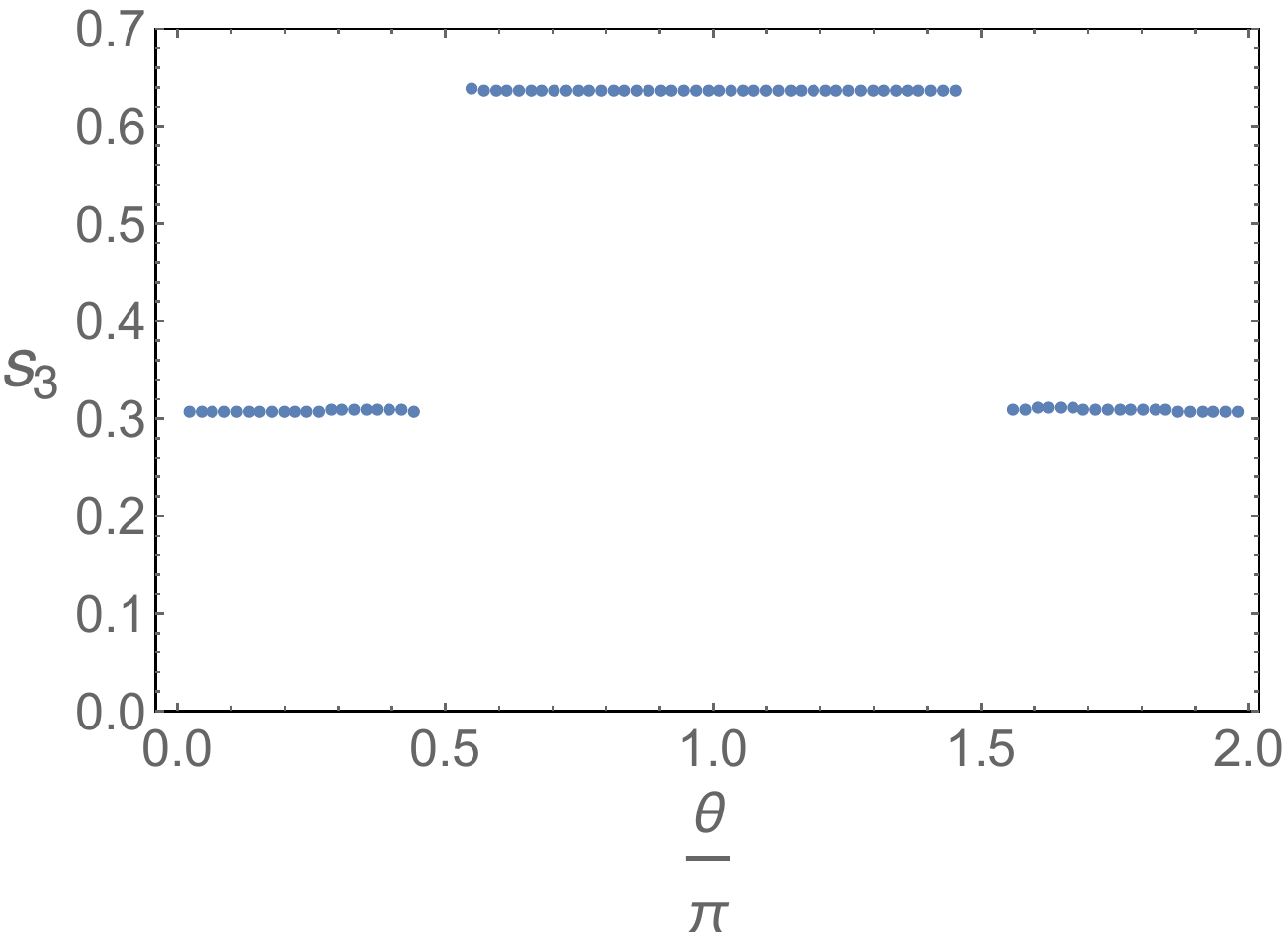}
\includegraphics[width=0.24\textwidth]{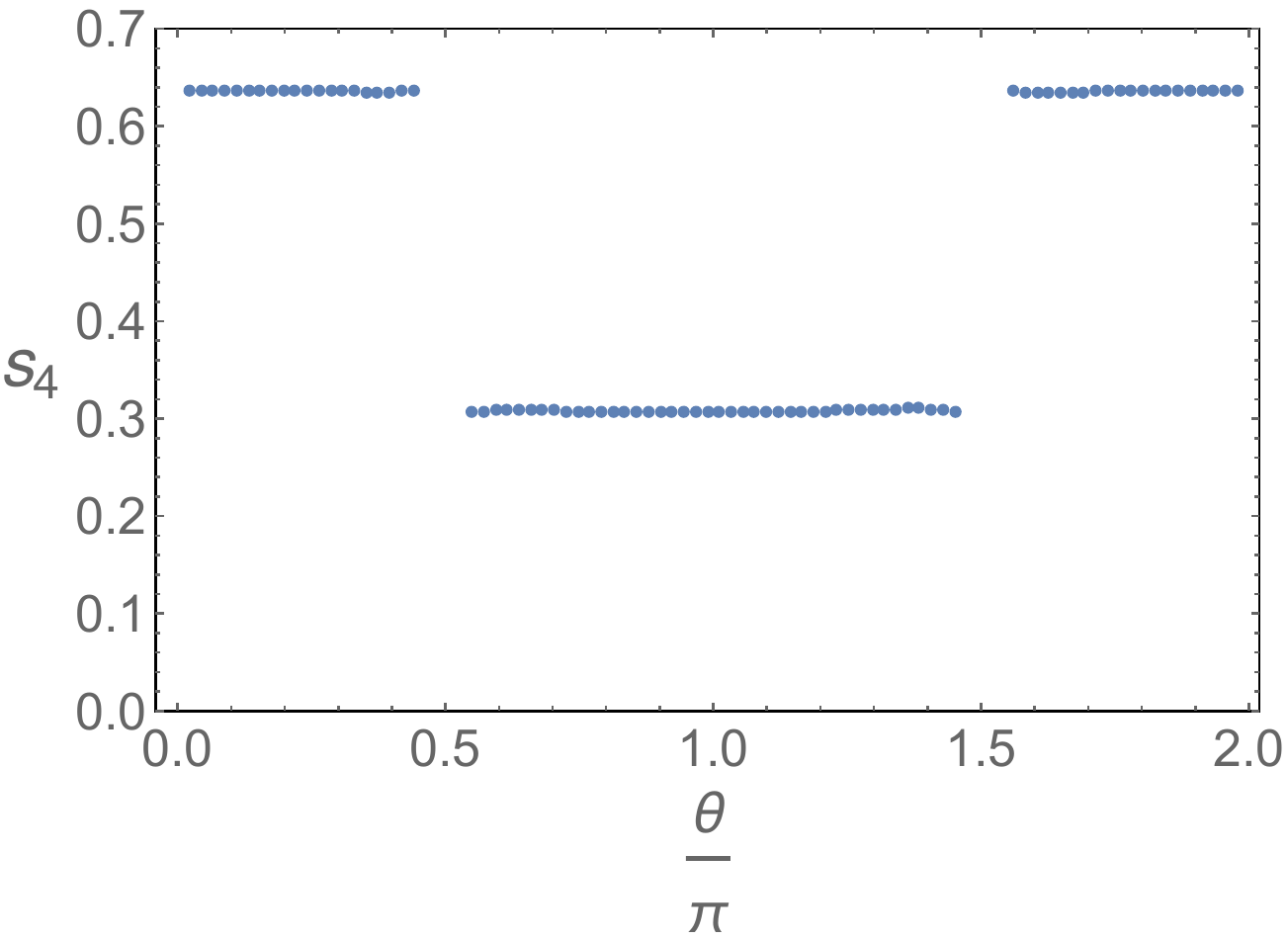}
  \caption{\small The eigenstates $(s_1, s_2, s_3, s_4)^T$ of $G^{-1}(0, \vec{k})$ as a function of $\theta/\pi$ along the enclosed loop with negative energy eigenvalues. Note that momentum is parameterized in (\ref{eq:discrete}). The sudden jump appears in all the four components of eigenstates $s_i$ across the $k_x$-axis i.e. at $\theta=\frac{1}{2}\pi,\frac{3}{2}\pi$.
  }
  \label{fig:eigenstate}
\end{figure}

When there is no sudden jump in the eigenstates, the normalized norm of the adjacent states in (\ref{eq:bpdis}) is one which contributes trivial to the phase factor of the Berry phase. The normalized norm at location of the sudden jump in the eigenstates is crucial for the phase factor which happens at $\theta=\pi/2, 3\pi/2$, similar to the case in the  weakly coupled field theory. We label the states close to this momentum according to Fig. \ref{fig:contribution}.

\begin{figure}[h!]
  \centering
  \includegraphics[width=0.450\textwidth]{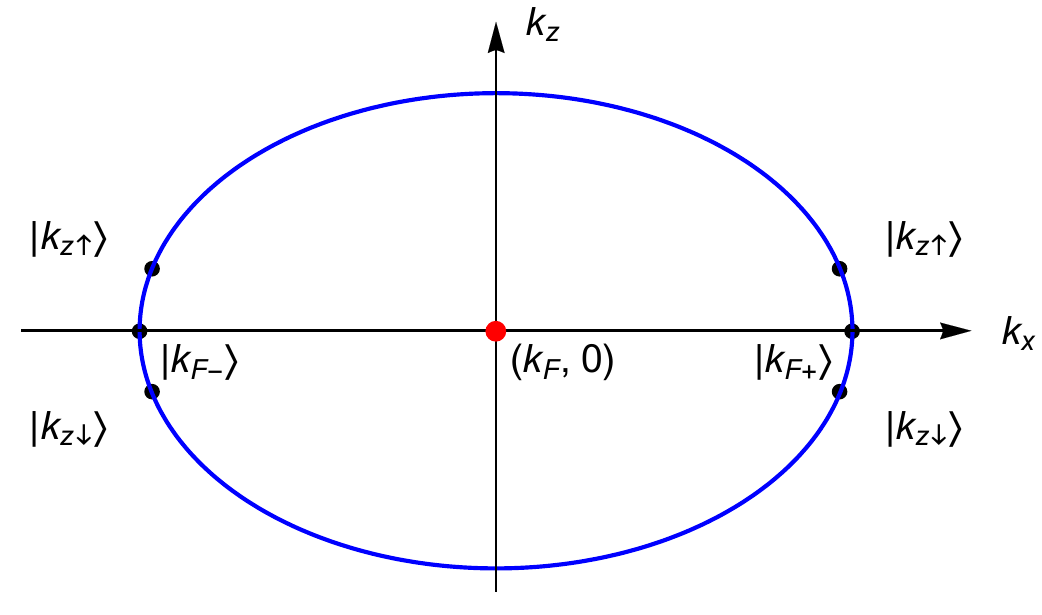}
  \caption{\small The diagram to illustrate the label of states which contributes nontrivially to the discrete Berry phase. The normalized inner product given by the adjacent points in the same upper or lower plane is one and gives a vanishing contribution to Berry phase. The net contribution of nontrivial discrete Berry phase is from the combination of $\langle k_{z\uparrow} | k_{F-}\rangle$, $\langle k_{F-} | k_{z\downarrow}\rangle$, $\langle k_{z\downarrow} | k_{F+}\rangle$ and $\langle k_{F+} | k_{z\uparrow}\rangle$, which can give a minus sign.}
  \label{fig:contribution}
\end{figure}

Close to these points $\theta=\pi/2, 3\pi/2$, the eigenvalues and the corresponding eigenstates are summarized as follows in table \ref{table:es}, where $b>a>0$. The values of $a, b$ are not important for analyzing the Berry phase.\footnote{The eigenstates are also quantitively related to the couplings $\eta_1, \eta_2$, but qualitatively the same as Fig. \ref{fig:eigenstate}. Therefore, the Berry phase and further the existence of the topological invariant does not rely on the choice of couplings.}
Substituting all these eigenstates into the discrete version of  Berry phase, only the inner products formed by the eigenstates in this table give out non-vanishing phase. 
Similar to the case of field theory, for each band we can compute the Berry phase for states with either positive or negative energy of  eigenvalues. 

With the states in table \ref{table:es} the calculation with respect to 
``Band-1" gives rise to a non-trivial phase $\pi$. 
However, the Berry phase for the nodal lines in ``Band-2" cannot be determined due to that the norm of the adjacent states vanishes close to $k_x=0$. 

All these features are qualitatively the same as those in the previous model \cite{Liu:2018bye}, indicating that 
the self-duality condition on the two form operators does not change the topological property in the strongly coupling regime. It seems that the topological property depends crucially on the action of the probe fermion while not the geometric background.\footnote{This is related to the fact that the probe fermionic operators are $1/N$ suppressed and their contributions to the geometric background is not visible at leading order.} This might indicate that there exist a semi-holographic description for the probe fermion similar to the case of probes with non-interacting Dirac equation \cite{Faulkner:2010tq}, from which the topological properties might be clearly shown. It would be extremely interesting to construct this semi-holographic description. 


\begin{table}[]
\centering
\renewcommand{\arraystretch}{1.5}
\begin{tabular}{|c|c|c|c|c|}
\hline
 &  $| k_{z\uparrow}\rangle$  &  $| k_{z\downarrow}\rangle$ & $|k_{F-}\rangle$  & $| k_{F+}\rangle$  \\  [1ex]\hline
\hline
 $E_{1}<0$ & $(b, -a, a, b)^T$ & $(a, -b, b, a)^T$ & $\frac{1}{2}(1, -1, 1, 1)^T$ & $\frac{1}{2}(1, 1, -1, 1)^T$  \\ [1ex]\hline
 $E_{1}>0$ & $(a, b, -b, a)^T$ & $(b, a, -a, b)^T$ & $\frac{1}{2}(1, 1, -1, 1)^T$ & $\frac{1}{2}(1, -1, 1, 1)^T$ \\ [1ex]\hline
\hline
 $E_{2}<0$ & $(b, -a, a, b)^T$ & $(a, -b, b, a)^T$ & $\frac{1}{2}(1, 1, 1, -1)^T$ & $\frac{1}{2}(1, -1, -1, -1)^T$  \\  [1ex]\hline
 $E_{2}>0$ & $(a, b, -b, a)^T$ & $(b, a, -a, b)^T$ & $\frac{1}{2}(1, -1, -1, -1)^T$ & $\frac{1}{2}(1, 1, 1, -1)^T$ \\  [1ex]\hline
\end{tabular}
\caption{\small We use $E_1$ and $E_2$ for eigenvalues of the states along discrete circles around ``Band-1" and ``Band-2". The states in the table are the ones which contribute to the nontrivial phase factor in the Berry phase. Note that here $b>a>0$.}
 \label{table:es} 
\end{table}
\section{Conclusion and discussion}
\label{sec:cd}
We have considered an improved holographic nodal line semimetal model in which the duality relation between the rank two operators $\bar{\psi}\gamma^{\mu \nu}\psi$ and $\bar{\psi}\gamma^{\mu \nu}\gamma^5\psi$ in the dual field theory is satisfied.  Following the approach in an AdS/QCD model \cite{Alvares:2011wb}, in holography we chose a special Chern-Simons term together with a mass term for the two form field to realize the duality constraint automatically.  
In this improved holographic nodal line semimetal, we found that 
there still exists a quantum phase transition from topological nodal line semimetal phase to the topologically trivial phase. 

We also calculated the fermionic spectral function by probing a massive Dirac fermion coupled in a particular way to the background field in bulk. 
We have found that, multiple nodal circle of Fermi surfaces 
exist in the topological NLSM phase while disappear in the trivial phase. The dispersion relation of low-energy excitations near the Fermi surfaces is linear. By tuning the dimensionless parameter $M/b$, the size of each nodal circle shrinks in the NLSM phase. 
These behaviors especially the discontinuity of Fermi circle in band structure indicate that a quantum phase transition happens at the critical point. 
We also computed the Berry phases in the NLSM phase and found that nodal circles are topologically nontrivial, which  confirms the topological property of NLSM phase and the broken of topology across the quantum phase transition, 

One might expect that after imposing the duality constraint new physics appears, while 
we have seen that most of the features above are qualitatively similar to the holographic model in \cite{Liu:2018bye}. 
The role of the duality in the two form operators does not modify the phase diagram qualitatively. The properties of the fermionic spectral function seems to depend crucially on the exact form the action of the probe fermions, while not on the geometric background. 
In condensed matter literature, the shear viscosity at low temperatures has been proposed in \cite{Moore:2019lay} as a probe to detect the band topology and the topological quantum phase transition.   Therefore, 
it would be interesting to study other physical quantities, like transports \cite{Moore:2019lay, Landsteiner:2016led, Ammon:2016fru, Bu:2018psl, Haack:2018ztx, Ammon:2020rvg} and non-local quantities to show some special features in this improved holographic model which serves as a natural ground to explore physics of strongly coupled NLSM.


\subsection*{Acknowledgments}
We would like to thank 
Elias Kiritsis, Karl Landsteiner and Ya-Wen Sun and especially Carlos Hoyos for useful discussions. 
This work is supported in part by the National Natural Science Foundation of China grant No.11875083.  

\appendix
\section{Equations of motion}
\label{appA}
The equations of motion from the action (\ref{eq:action}) in the main text are
\bea
\label{eq:ein1}
\begin{split}
R_{ab}-\frac{1}{2}g_{ab}(R+12)-T_{ab}&=0\,,\\
\nabla_b \mathcal{F}^{ba}+2\alpha \epsilon^{abcde} F_{bc}\mathcal{F}_{de}&=0\,,\\
\nabla_b F^{ba}+\alpha \epsilon^{abcde} (F_{bc}F_{de}+\mathcal{F}_{bc}\mathcal{F}_{de})
~~~~~~&\\
-iq_1\big(\Phi^*D^a\Phi-(D^a\Phi)^*\Phi\big)
+\frac{q_2}{\eta}\epsilon^{abcde} B_{bc}B_{de}^*&=0\,,\\
D_a D^a\Phi-\partial_{\Phi^*} V_1-\lambda\Phi B_{ab}^*B^{ab}&=0\,,\\
\frac{i}{3\eta} \epsilon_{abcde} H^{cde}-m_2^2 B_{ab}-\lambda \Phi^*\Phi B_{ab}&=0\,,
\end{split}
\eea
where 
\bea
\begin{split}
T_{ab}&=\frac{1}{2}\Big[\mathcal{F}_{ac}\mathcal{F}_{b}^{~c}-\frac{1}{4}g_{ab}\mathcal{F}^2\Big]+
\frac{1}{2}\Big[F_{ac} F_{b}^{~c}-\frac{1}{4}g_{ab}F^2\Big]+\frac{1}{2}\big((D_a\Phi)^*D_b\Phi+(D_b\Phi)^*D_a\Phi\big)\\
&+(m_2^2+\lambda|\Phi|^2)(B_{ac}^*B_b^{~c}+B_{bc}^*B_a^{~c})-\frac{1}{2}\Big((D_c \Phi)^*(D^c\Phi)+V_1+V_2+\lambda|\Phi|^2B_{cd}^*B^{cd} \Big)g_{ab}\,
\end{split}
\eea
is the energy-momentum tensor.
We make use of the following ansatz for the zero-temperature solutions, 
\bea
\label{eq:bg-nlsm}
\begin{split}
ds^2 &= u(-dt^2+dz^2)+\frac{dr^2}{u}+f(dx^2+dy^2)\,,\\~~~
\Phi&=\phi(r)\,,~~~\\
B_{xy}&=-B_{yx}=\mathcal{B}_{xy}\,,\\
B_{tz}&=-B_{zt}=i\mathcal{B}_{tz}\,.
\end{split}
\eea
The equations of motion can be explicitly written as
\bea
\label{eq:background}
\begin{split}
\frac{u''}{u}+\frac{f''}{f}+\frac{1}{3}\left(\frac{u'^2}{2u^2}+\frac{7u'f'}{2uf}-\frac{f'^2}{f^2} \right)-\frac{8}{u}+\frac{2}{3}\phi'^2+\frac{2}{3u}\left(m_1^2\phi^2+\frac{\lambda_1}{2}\phi^4\right)&=0\,\\
\frac{u''}{u}-\frac{f''}{f}+\frac{u'}{2u}\left(\frac{u'}{u}-\frac{f'}{f}\right)-4(m_2^2+\lambda\phi^2)\left(\frac{\mathcal{B}_{tz}^2}{u^3}+\frac{\mathcal{B}_{xy}^2}{uf^2}  \right) &=0\,\\
\phi''+\left(\frac{3u'}{2u}+\frac{f'}{f}\right)\phi'-\left(m_1^2+\lambda_1\phi^2-\frac{2\lambda\mathcal{B}_{tz}^2}{u^2}+\frac{2\lambda \mathcal{B}_{xy}^2}{f^2} \right)\frac{\phi}{u}&=0\,\\
\mathcal{B}_{tz}'-\frac{\eta \sqrt{u}}{2f}(m_2^2+\lambda \phi^2)\mathcal{B}_{xy}&=0\,\\
\mathcal{B}_{xy}'-\frac{\eta f}{2u^{\frac{3}{2}}}(m_2^2+\lambda \phi^2)\mathcal{B}_{tz}&=0\,\\
\end{split}
\eea
There is an extra first order constraint equation which can be expressed as linear combinations of the previous equations and their derivatives
\be
\label{eq:constraint}
\frac{\phi'^2}{2}-\left(\frac{u'^2}{4u^2}+\frac{f'^2}{4f^2}+\frac{u'f'}{uf}\right)+\frac{6}{u}-\frac{1}{2u}\left(m_1^2\phi^2+\frac{1}{2}\lambda_1\phi^4 \right)
+\left(\frac{m_2^2+\lambda \phi^2}{u}\right)\left(\frac{\mathcal{B}_{tz}^2}{u^2}-\frac{\mathcal{B}_{xy}^2}{f^2} \right)=0\,.
\ee

\section{Counterterms and on-shell action}
\label{appB}
To make the gravitational theory well behaved in variation and remove the divergence in the on-shell action, the Gibbons-Hawking term $S_{\text{GH}}$ and the counterterms $S_{\text{c.t}}$ should be considered to construct the renormalized action 
\bea
S_{\text{ren}}=S + S_{\text{GH}}+S_{\text{c.t}}\,,
\eea
where 
\bea
\begin{split}
S_{\text{GH}}&=\int _{\partial}d^dx \sqrt{-h}\left(2K\right)\,,\\
S_{\text{c.t}}&=\int_{\partial}d^dx \sqrt{-h}\left[-6-\Phi^2+\frac{1}{2}|B_{\mu\nu}|^2+\text{ln}~r\left((\frac{1}{3}+\frac{\lambda_1}{2})\Phi^4+|B_{\mu\nu}|^4\right)\right]
\end{split}
\eea
are defined on the boundary of the bulk. $K=h^{ab}\nabla_a n_b$ is the trace of the extrinsic curvature of the induced metric $h_{ab}=g_{ab}-n_a n_b$ with $n_a=(0, 0, 0, 0, \frac{1}{\sqrt{u}})$ the spacelike normal vector, and $h$ is the determinant of the induced metric reduced onto the hypersurface orthogonal to $n_{a}$, i.e., $h\equiv \text{det}~h_{\mu\nu(\mu, \nu\neq r)}$.

Close to the AdS boundary $(r\rightarrow \infty)$, the expansions of the fields are
\bea
\label{eq:bguv}
\begin{split}
u\big{|}_{r\rightarrow \infty} &=r^2-2b^2-\frac{M^2}{3}+\left(\frac{4b^4}{9}+\frac{23M^4}{180}\right)\frac{\text{ln}(r)}{r^2}+\frac{u_2}{r^2}+...\,\\
f\big{|}_{r\rightarrow \infty} &=r^2-\frac{M^2}{3}+\left(\frac{4b^4}{9}+\frac{23M^4}{180}\right)\frac{\text{ln}(r)}{r^2}+\frac{f_2}{r^2}+...\,\\
\phi\big{|}_{r\rightarrow \infty} &=\frac{M}{r}-\frac{23M^3}{60}\frac{\text{ln}(r)}{r^3}+\frac{\phi_2}{r^3}+...\,\\
\mathcal{B}_{tz}\big{|}_{r\rightarrow \infty} &=br-2b^3\,\frac{\text{ln}(r)}{r}+\frac{b_{tz2}}{r}+...\,\\
\mathcal{B}_{xy}\big{|}_{r\rightarrow \infty} &=br+2b^3\,\frac{\text{ln}(r)}{r}+\frac{b_{xy2}}{r}+...\,\\
\end{split}
\eea
where 
\bea
\begin{split}
f_2&=\frac{7b^4}{18}+\frac{b b_{xy2}}{3}+\frac{5b^2M^2}{18}+\frac{149M^4}{1440}-\frac{u_2}{2}-\frac{M\phi_2}{2}\,\\
b_{tz2}&=-b^3-b_{xy2}-\frac{7 b M^2}{6}\,\\
\end{split}
\eea
together with $\{b, M, u_2, b_{xy2}, \phi_2\}$ are the coefficients of the series expansions. 
We have numerically checked that when $b=1$ is fixed, 
all the 
coefficients of the series expansions change smoothly by tuning 
$M$, even pass across the critical point. 
The free energy density is $\frac{\Omega}{V}=-\frac{S_{\text{o.s}}}{V}$ can be expressed by these 
coefficients 
\bea
S_{\text{o.s}}=\frac{1}{b^4}\left(\frac{11b^4}{9}-\frac{8 b b_{xy2}}{3}-\frac{38b^2M^2}{9}-\frac{7M^4}{36}+3u_2+2M\phi_2\right), 
\eea
which means that the free energy density is also smooth through the phase transition as illustrated in Fig. \ref{fig:freeE} in the main text. 

\section{Scaling symmetries and numerical calculation}
\label{appC}
The following scaling symmetries are very useful 
for numerical calculations. 
\begin{itemize}
\item $\{r^{-1}, t, x, y, z\}\rightarrow \{\tilde{r}^{-1}, \tilde{t}, \tilde{x}, \tilde{y}, \tilde{z}\}=b\{r^{-1}, t, x, y, z\}$, while $\{u, f, B_{\mu\nu}\} \rightarrow \{\tilde{u}, \tilde{f}, \tilde{B}_{\mu\nu}\}=b^{-2}\{u, f, B_{\mu\nu}\}$ to make $ds^2$ and $B=B_{\mu\nu}dx^{\mu}dx^{\nu}$ remain unchanged according to this transformation. This symmetry can be used to fix $b$ to be 1. 
\item
$\{x, y\}\rightarrow \{\tilde{x}, \tilde{y}\}=c\{x, y\}$ together with $\{f, B_{xy}\} \rightarrow \{\tilde{f}, \tilde{B}_{xy}\}=c^{-2}\{f, B_{xy}\}$ indicates another scaling symmetry that is restricted in the $x$-$y$ plane. 
This symmetry can scale $f$ to asymptotic to $r^2$ near the boundary. It also makes us possible to fix a shooting parameter in the IR region, since we can make a transformation back to the expected coordinates.

\end{itemize}

\section{Dirac system in the bulk}
\label{appD}
\subsection{Vielbein and spin connection}
The vielbein is a two indexed object with a tangent space index $\underline{m}$, and a coordinate index $a$. It obeys the relations $e^a_{\underline{m}}e^b_{\underline{n}} g_{ab}=\eta_{\underline{m}\underline{n}}$ and $\eta_{\underline{m}\underline{n}} e^{\underline{m}}_ae^{\underline{n}}_b=g_{ab}$.
For a diagonal metric (\ref{eq:bg-nlsm}), we have 
\bea
e^{\underline{m}}_a=\sqrt{|g_{aa}|}\delta^{\underline{m}}_a
\eea
where $a$ does not sum. The tangent space index $\underline{m}$ is lowered or raised by Minkowski metric $\eta_{\underline{m}\underline{n}}$ or $\eta^{\underline{m}\underline{n}}$, while the coordinate index is lowered or raised by $g_{ab}$ or $g^{ab}$. 

The spin connection can be constructed from the vielbein and the Christoffel symbol 
\bea
\omega^{~\underline{m}}_{a~~\underline{n}}=\Gamma^b_{~ac} e^{\underline{m}}_b e^c_{\underline{n}}-e^b_{\underline{n}}\partial_a e^{\underline{m}}_b\,.
\eea
The covariant derivative of the spinor can be defined as 
\bea
\nabla_a=\partial_a-\frac{i}{4}\omega_{a\underline{m}\underline{n}}\Gamma^{\underline{m}\underline{n}}\,.
\eea

\subsection{Gamma matrices and spinors}

We use the Pauli matrices
\bea
\label{eq:Pauli}
\sigma_x=
\left(                
  \begin{array}{cc}   
   0 &~ 1\\ 
   1 &~ 0\\
  \end{array}
\right)\,,~~       
\sigma_y=
\left(                
  \begin{array}{cc}   
   0 &~ -i\\ 
   i &~ 0\\
  \end{array}
\right)\,,~~
\sigma_z=
\left(                
  \begin{array}{cc}   
   1 &~ 0\\ 
   0 &~ -1\\
  \end{array}
\right)\,,~~
I_2=
\left(                
  \begin{array}{cc}   
   1 &~ 0\\ 
   0 &~ 1\\
  \end{array}
\right)\,.   
\eea
to build up the gamma-matrices $\gamma^\mu$ in the 4-dimensional flat space-time
\bea
\label{eq:gamma-4D}
\gamma^0=
\left(                
  \begin{array}{cc}   
   0 &~ i I_2\\ 
   i I_2 &~ 0\\
  \end{array}
\right)\,, ~~   
\gamma^i=
\left(                
  \begin{array}{cc}   
   0 &~i  \sigma_i\\ 
   -i \sigma_i &~ 0\\
  \end{array}
\right)\,, ~~ 
\gamma^5=i\gamma^0\gamma^1\gamma^2\gamma^3=
\left(                
  \begin{array}{cc}   
   -I_2 &~ 0\\ 
   0 &~ I_2\\
  \end{array}
\right)\,.
\eea
The gamma matrices $\Gamma^{\underline{a}}$ in the 5-dimensional local flat space-time can be constructed 
\bea
\label{eq:gamma-5D}
\Gamma^{\underline{a}}=\left(\Gamma^{\underline{\mu}},~\Gamma^{\underline{r}}\right) \equiv \left(\gamma^\mu, -\gamma^5\right)\,.
\eea
Gamma matrices $\Gamma^a$ in the 5-dimensional curved space-time are 
\bea
\Gamma^a=e^a_{\underline{m}}\Gamma^{\underline{m}}\,.
\eea
The Clifford algebras for the above cases are
\bea
\label{eq:Clifford}
\{\gamma^\mu, \gamma^\nu\}=2\eta^{\mu\nu}\,, ~~~\{\Gamma^{\underline{a}}, \Gamma^{\underline{b}}\}=2\eta^{\underline{ab}}\,, ~~~\{\Gamma^a, \Gamma^b\}=2g^{ab}\,.
\eea
The two-indexed anti-symmetrized products of gamma matrices are defined as
\bea
\gamma^{\mu\nu}=\frac{i}{2}\left[\gamma^\mu, \gamma^\nu\right], ~~\Gamma^{\underline{ab}}=\frac{i}{2}\left[\Gamma^{\underline{a}}, \Gamma^{\underline{b}}\right],~~\Gamma^{ab}=\frac{i}{2}\left[\Gamma^a, \Gamma^b\right]=e^a_{\underline{m}}e^b_{\underline{n}}\Gamma^{\underline{mn}}\,.
\eea

A spinor $\psi$ can be decomposed into the right-handed and left-handed spinors $\psi_{R,L}$ defined as
\bea
\psi_R=\left(
  \begin{array}{c}   
   \psi_+\\ 
   0\\
  \end{array}
\right),~~~
\psi_L=\left(
  \begin{array}{c}   
   0\\ 
   \psi_-\\
  \end{array}
\right)
\eea
with the projection operator constructed from $\Gamma^{\underline{r}}$
\bea
\frac{1}{2}(1\pm \Gamma^{\underline{r}})\psi=\psi_{R,L}
\eea
where $\psi_{\pm}$ are two-components spinors.

\subsection{UV boundary terms of Dirac equations}

In this part, we briefly review 
how to obtain the correct boundary action for a single chiral fermion at the boundary \cite{Henneaux:1998ch, Iqbal:2009fd, Laia:2011zn}. Next, we review the generalization to combining two opposite chiral fermions into a massive Dirac fermion with correct boundary action. 

We start from the 
action for a single, free Dirac fermion with mass $m$ in the bulk
\bea
S
_{\text{Dirac}}=\int d^{d+1}x\sqrt{-g}\left(\bar{\psi}\Gamma^a \nabla_a\psi-m\bar{\psi}\psi\right)\,. 
\eea
Note that the boundary is defined at $r\to\infty$. 
The variation of the action is 
\bea
\delta S_{\text{Dirac}}=\text{bulk term}+\int_{\partial \mathcal{M}}d^dx\sqrt{-gg^{rr}}\left(\bar{\psi}_L\delta\psi_R-\bar{\psi}_R\delta\psi_L\right)
\eea
where the bulk terms give the dynamical equations of motion. The last terms are located at the AdS boundary 
where $\psi_R$ and $\psi_L$ as 
varied independently. 
However, because the bulk Dirac equation is of first order, 
$\psi_R$ and $\psi_L$ are related and only one of them can be varied freely while the other behaves as the corresponding response. This can be achieved 
by adding a proper boundary term to the original 
action \cite{Henneaux:1998ch, Iqbal:2009fd, Laia:2011zn}. 

For example, if we choose $\psi_R$ as free variable, we add
\bea
\label{Eq:Sbd1}
S_{\partial}=\int_{\partial\mathcal{M}}d^dx\sqrt{-gg^{rr}}\bar{\psi}_R\psi_L\,,
\eea
and 
the variation of the boundary field theory is 
\bea
\label{Eq:var1}
\delta S_{\text{total}}=\delta(S_{\text{bulk}}+S_{\partial})=\int_{\partial \mathcal{M}}d^dx\sqrt{-gg^{rr}}\left(\bar{\psi}_L\delta\psi_R+\delta\bar{\psi}_R\psi_L\right)\,.
\eea
Alternatively, 
we can take $\psi_L$ as the free variable by considering the  boundary term
\bea
\label{Eq:Sbd2}
S_{\partial}=\int_{\partial \mathcal{M}}d^dx\sqrt{-gg^{rr}}\left(-\bar{\psi}_L\psi_R\right),
\eea
leading to the alternative total variation as
\bea
\label{Eq:var2}
\delta S_{\text{total}}=\delta(S_{\text{bulk}}+S_{\partial})=-\int_{\partial \mathcal{M}} d^dx\sqrt{-gg^{rr}}\left(\delta \bar{\psi}_L\psi_R+\bar{\psi}_R\delta \psi_L \right)\,.
\eea

For five dimensional bulk theory, we introduced two sets of coupled fermions $\psi^{(1)}, \psi^{(2)}$ in (\ref{eq:bulkdirac5d}). 
The boundary terms we considered are 
a combination of (\ref{Eq:Sbd1}) and  (\ref{Eq:Sbd2}).  
Performing the variation, 
we end up with 
\bea
S_{\text{bdy}}= \int d^dx\sqrt{-h}\left(\bar{\psi}^{(1)}_R\psi_L^{(1)}-\bar{\psi}_L^{(2)}\psi_R^{(2)}\right)
\eea
where the mass of $\psi^{(1)}$ and $\psi^{(2)}$ are $m$ and $-m$, respectively. 
One may also notice that there is an opposite sign comparing  (\ref{Eq:var1}) and (\ref{Eq:var2}). This results in the additional minus sign in (\ref{eq: sourceresponse}) when identifying the 4-component fermionic operator.



\subsection{IR boundary conditions}
\label{appD4}
We outline the infalling boundary conditions according to different phases and choices of $k_{\mu}=(\omega, k_x, k_y, k_z)$ in a table.
\begin{table}[H]
\centering
\renewcommand{\arraystretch}{1.5}
\begin{tabular}{|c|c|c|}
\hline
 Phase &  $k^\mu$  & IR infalling solution\\ \hline
\multirow{2}{*}{NLSM} & $\omega$ or $k_z \neq 0, \forall ~k_x, k_y$ & $\psi^{\text{IR}}_l=e^{i\frac{\sqrt{\omega^2-k_z^2}}{u_0 r}}\left(z_1^l\,,~ z_2^l\,,~ i\frac{\sqrt{\omega^2-k_z^2}}{\omega-k_z}z_1^l\,,~ i\frac{\sqrt{\omega^2-k_z^2}}{\omega+k_z}z_2^l\right)^T$ \\[1ex] \cline{2-3}
 & $\omega=k_z=0, k_x$ or $k_y\neq 0$ &$\psi^{\text{IR}}_l=e^{-\frac{2\tilde{k}_x}{\alpha}r^{-\frac{\alpha}{2}}}\left(z_1^l\,, ~z_2^l\,,~ z_2^l\,,~ z_1^l\right)^T$\\[1ex] \hline
 Critical &  $\omega=k_z=0, k_x$ or $k_y\neq 0$  & $\psi^{\text{IR}}_l=e^{-\frac{2\tilde{k}_x}{\alpha_c}r^{-\frac{\alpha_c}{2}}}\left(z_1^l\,, ~z_2^l\,,~ z_2^l\,,~ z_1^l\right)^T$\\[1ex] \hline
 Trivial &  $\omega=k_z=0, k_x$ or $k_y\neq 0$  & $\psi^{\text{IR}}_l=e^{-\frac{\tilde{k}_x}{r}}\left(z_1^l\,, ~z_2^l\,,~ z_2^l\,,~ z_1^l\right)^T$\\[1ex] \hline
\end{tabular}
\end{table}
In this table $\tilde{k}_x\equiv \frac{k_x}{\sqrt{u_0 f_0}}$ in NLSM phase, $\tilde{k}_x\equiv \frac{k_x}{\sqrt{u_c f_c}}$ at the critical point and $\tilde{k}_x\equiv \frac{k_x}{\sqrt{u_1}}$, 
with $(u_0, f_0, u_c, f_c, u_1)=\left(\frac{1}{8}(11 + 3\sqrt{13}), \sqrt{\frac{2}{3}\sqrt{13}-2}, 2.735, 0.754, 1+\frac{3}{8\lambda_1}\right)$.




 
\end{document}